\documentclass[useAMS,usenatbib]{mn2e}
\usepackage{txfonts}
\usepackage{graphicx} 
\usepackage{rotating}

\DeclareRobustCommand*\cal{\@fontswitch\relax\mathcal}

\title[A Bayesian blind survey]{A Bayesian blind survey for cold molecular gas in the Universe}
\author[L. Lentati, C. Carilli, P. Alexander, F. Walter, R. Decarli]{L. Lentati$^{1}$\thanks{E-mail:
ltl21@cam.ac.uk} C. Carilli$^{2,1}$ P. Alexander$^{1}$ F. Walter$^{3}$ R. Decarli$^{4}$\\
$^{1}$Astrophysics Group, Cavendish Laboratory, JJ Thomson Avenue,  Cambridge, CB3 0HE, UK\\
$^{2}$NRAO, Pete V. Domenici Array Science Center, P.O. Box O, Socorro, NM, 87801, USA \\
$^{3}$Max-Planck Institut fur Astronomie, Konigstuhl 17, D-69117, Heidelberg, Germany \\
$^{4}$Laboratoire AIM, CEA/DSM-CNRS-Universite Paris Diderot, Irfu/Service d'Astrophysique, CEA Saclay, Orme des Merisiers, 91191 Gif-sur-Yvette cedex, France}
\begin{document}

\maketitle

\label{firstpage}

\begin{abstract}
A new Bayesian method for performing an image domain search for line-emitting galaxies is presented. The method uses both spatial and spectral information to robustly determine the source properties, employing either simple Gaussian, or other physically motivated models whilst using the evidence to determine the probability that the source is real.   In this paper, we describe the method, and its application to both a simulated data set, and a blind survey for cold molecular gas using observations of the Hubble Deep Field North taken with the Plateau de Bure Interferometer.  We make a total of 6 robust detections in the survey, 5 of which have counterparts in other observing bands. We identify  the most secure detections found in a previous investigation, while finding one new probable line source with an optical ID not seen in the previous analysis.  This study acts as a pilot application of Bayesian statistics to future searches to be carried out both for low-$J$ CO transitions of high redshift galaxies using the JVLA, and at millimeter wavelengths with ALMA, enabling the inference of robust scientific conclusions about the history of the molecular gas properties of star-forming galaxies in the Universe through cosmic time. 
\end{abstract}


\section{Introduction}

The systematic detection and parameterization of sources of interest in astronomical images is central to a vast array of different scientific goals.  These can range from exploring the nature of galaxy formation using high resolution imaging of clumpy, extended gas resevoirs (e.g. \citealt{2012ApJ...760...11H}), to surveys such as BOSS (e.g. \citealt{2013AJ....145...10D}), that seek to characterise barycentric acoustic oscillations (the imprint of acoustic waves that propogated pre-recombination on the large scale structure of the Universe).

A common tool in source finding is the SERCH algorithm described in \cite{1991ApJ...377L..65U}.  In brief, a Gaussian kernal of specified width is used to perform a weighted average over a set of channels, after which detections over some signal to noise (S:N) threshold will be accepted as candidate objects.  Different kernal widths can then be iterated over such that the maximum S:N can be obtained for each candidate.  Ultimately however this approach relys on the user to determine via visual inspection whether or not to consider a candidate real, or simply a fluctuation in the noise.  This is a problem made all the worse given any real image will likely contain artifacts and spikes in the noise that will sum coherently with the signal using such an approach, making manual post processing a necessity. 

More recently the DUCHAMP algorithm  \citep{2012MNRAS.421.3242W} was introduced that does not impose any spectral or spatial model for the source during the search and so returns only the locations of candidates above some S:N threshold.  It allows the user to smooth the image either spatially or spectrally at some chosen scale in order to improve S:N at that scale, or uses a wavelet reconstruction that does not require any assumptions be made about the size of the objects in the image, but assumes that the noise between pixels is uncorrelated, which will not be true for any interferometric observation where the point spread function (PSF) of the interferometer acts to correlate the noise across the image.  

In \cite{2014ApJ...782...78D}, henceforth D14, an analysis is described of a blind survey of the Hubble Deep Field North covering the entire 3mm window (79-115~GHz) using the IRAM Plateau de Bure Interferometer (PdBI).  Here two different techniques were introduced in order to perform the search.  The first defines a `spread' parameter that models the deviation in the noise from a normal Gaussian distribution introduced by the prescense of a source due to an excess of positive flux.  The second is based on the finder `cprops' which uses a simple signal to noise cut of the image cube, onto which a mask is applied that identifies an excess over multiple concurrent channels.  The lack of any parametric model being applied to the properties of the source, however, both in these methods, and in the DUCHAMP algorithm, makes it difficult to objectively quantify how likely that source is to be real given our knowledge of galaxy formation.  The preferred method should be to take a set of physically meaningful models and to use these in the search, determining in a robust way the probability that any given model is supported by the data. 

In this paper we present a Bayesian search algorithm designed to address just such problems.  We fit different parametric models to an observed dataset using the inference tool MULTINEST \citep{2008MNRAS.384..449F, 2009MNRAS.398.1601F} to efficiently calculate the Bayesian evidence and so objectively quantify the probability of the existence of those sources, and allow for selection between the different models.

Whilst this algorithm has already been used in several publications (e.g. \cite{2012MNRAS.426..258A, 2013ApJ...773...44W,2012Natur.486..233W}) here we describe the details of the method, and apply it to both a simulated data set, and to the PdBI survey of the Hubble Deep Field North.  Details of the survey and its cosmological implications are presented in detail in D14 and \cite{2014ApJ...782...79W} (henceforth W14).  Here we focus on the technical aspects of our methodology of performing such a search and the results of its application to this survey.

In sections \ref{Section:Bayes} and \ref{Section:mu} we discuss Bayes' theory, and its application to our search method in terms of quantifying the proability that a source exists in the data.  In section \ref{section:likes} we discuss the technical aspects of the algorithm, and the galactic models used.  Section \ref{Section:Sim} sees this method applied to a simulated data set designed to represent the proceeding survey, and finally in section \ref{Section:hdf} we show our results of performing the search on the PdBI observation with concluding remarks in section \ref{section:conclusions}.  

We adopt a concordance $\Lambda$CDM cosmology throughout, with $H_0$ = 71~km$\mathrm{s^{-1}Mpc^{-1}}$, $\Omega_{\mathrm{M}}$ = 0.27, and $\Omega_{\mathrm{\Lambda}}$ = 0.73 \citep{2007ApJS..170..377S}.

\section{Bayesian Inference} 
\label{Section:Bayes}

Our galaxy modeling methodology is built upon the principles of Bayesian inference; here we give a summary of this framework.  Bayesian inference methods provide a consistent approach to the estimation of a set of parameters $\mathbf{\Theta}$ in a model or hypothesis $\mathcal{H}$ given a set of data, $\mathbf{D}$.  Bayes' theorem states that

\begin{equation}
\mathrm{Pr}(\mathbf{\Theta} \mid \mathbf{D}, \mathcal{H}) = \frac{\mathrm{Pr}(\mathbf{D}\mid \mathbf{\Theta}, \mathcal{H})\mathrm{Pr}(\mathbf{\Theta} \mid \mathcal{H})}{\mathrm{Pr}(\mathbf{D} \mid \mathcal{H})},
\end{equation}
where $\mathrm{Pr}(\mathbf{\Theta} \mid \mathbf{D}, \mathcal{H}) \equiv \mathrm{Pr}(\mathbf{\Theta})$ is the posterior probability distribution of the parameters,  $\mathrm{Pr}(\mathbf{D}\mid \mathbf{\Theta}, \mathcal{H}) \equiv L(\mathbf{\Theta})$ is the likelihood, $\mathrm{Pr}(\mathbf{\Theta} \mid \mathcal{H}) \equiv \pi(\mathbf{\Theta})$ is the prior probabiltiy of the parameters given the model, and $\mathrm{Pr}(\mathbf{D} \mid \mathcal{H}) \equiv Z$ is the Bayesian Evidence.  

In parameter estimation, the normalizing evidence factor is usually ignored, since it is independent of the parameters $\mathbf{\Theta}$, and inferences are obtained by taking samples from the (unnormalised) posterior using, for example, standard Markov chain Monte Carlo (MCMC) sampling methods.  The posterior obtained constitutes the complete Bayesian inference of the parameter values, and can, for example, be marginalized over each parameter to obtain individual parameter constraints.

In contrast to parameter estimation, for model selection the evidence takes the central role and is simply the factor required to normalize the posterior over $\mathbf{\Theta}$:

\begin{equation}
Z = \int L(\mathbf{\Theta})\pi(\mathbf{\Theta}) \mathrm{d}^n\mathbf{\Theta},
\label{eq:Evidence}
\end{equation}
where $n$ is the dimensionality of the parameter space.  As the average of the likelihood over the prior, the evidence is larger for a model if more of its parameter space is likely and smaller for a model with large areas in its parameter space having low likelihood values, even if the likelihood function is very highly peaked.  Thus, the evidence automatically implements Occam's razor: a simpler theory with a compact parameter space will have a larger evidence than a more complicated one, unless the latter is significantly better at explaining the data.  The question of model selection between two models $\mathcal{H}_0$ and $\mathcal{H}_1$ can then be decided by comparing their respective posterior probabilities, given the observed data set $\mathbf{D}$, via the model selection ratio $R$:

\begin{equation}
R= \frac{\mathrm{Pr}(\mathcal{H}_1\mid \mathbf{D})}{\mathrm{Pr}(\mathcal{H}_0\mid \mathbf{D})} = \frac{\mathrm{Pr}(\mathbf{D} \mid \mathcal{H}_1)\mathrm{Pr}(\mathcal{H}_1)}{\mathrm{Pr}(\mathbf{D}\mid \mathcal{H}_0)\mathrm{Pr}(\mathcal{H}_0)} = \frac{Z_1}{Z_0}\frac{\mathrm{Pr}(\mathcal{H}_1)}{\mathrm{Pr}(\mathcal{H}_0)},
\label{Eq:Rval}
\end{equation}
where $\mathrm{Pr}(\mathcal{H}_1)/\mathrm{Pr}(\mathcal{H}_0)$ is the a priori probability ratio for the two models, which can often be set to unity but occasionally requires further consideration.

Evaluation of the multidimensional integral in \ref{eq:Evidence} is a challenging numerical task.  The nested sampling approach, introduced by Skilling (2004), is a Monte-Carlo method targeted at the efficient calculation of the evidence, but also produces posterior inferences as a by-product.  Feroz $\&$ Hobson (2008) and Feroz et al. (2008) built on this nested sampling framework, and introduced the MULTINEST algorithm, which is very efficient in sampling from posteriors that may contain multiple modes and/or large (curving) degeneracies, and also calculates the evidence.  This technique has greatly reduced the computational cost of Bayesian parameter estimation and model selection, and is employed in this paper.

\section{Quantification of source detection}
\label{Section:mu}

We now discuss how one may calculate the probability that the observed field does indeed contain a real galaxy.  This quantification is most naturally performed via a Bayesian model selection by evaluating the evidence associated with the posterior for competing models for the data (see e.g. Hobson $\&$ McLachlan (2003)). It is convenient to consider the following models:
\begin{itemize}
\item$\mathcal{H}_{\mathrm{0}}$ = No galaxy exists in $S$

\item$\mathcal{H}_{\mathrm{1}}$ = A galaxy exists in $S$
\end{itemize}
where $S$ is the spatial volume contained in the prior.
We must calculate the model selection ratio $R$ given in Eq. \ref{Eq:Rval} between the hypotheses $\mathcal{H}_i$.  For each hypothesis $\mathcal{H}_i (i=0,1)$, the evidence is given by

\begin{equation}
Z_i = \int L(\mathbf{\Theta})\pi_i(\mathbf{\Theta}) ~\mathrm{d}^n\mathbf{\Theta},
\label{Eq:evidences}
\end{equation}
where

\begin{equation}
\pi_i(\mathbf{\Theta}) = \prod_{p=1}^{N_{\mathrm{dim}}}\pi_i(\theta_p),
\end{equation}
is the prior for the $N_{\mathrm{dim}}$ parameters that describe our model for hypothesis $i$, and $\pi_i(\theta_p)$ the prior for parameter $p$.  The priors on all model parameters apart from the amplitude of the source $A$ may be taken to be the same in each hypothesis.  For source amplitude however, we take $\pi_1(A)$ to be uniform between some range $[0, A_{\mathrm{max}}]$ and $\pi_0(A)$ is a delta function centered on A=0.

Therefore our evidence for $\mathcal{H}_0$ will be:

\begin{equation}
Z_0 = \frac{1}{|S|}\int\;L(\mathbf{X}, A=0, R) ~\mathrm{d}\mathbf{X} = L(\mathbf{X}, A=0, R) \equiv L_0,
\end{equation}
which is independant of both $S$ and the particular set of parameters $\Theta$.  

Following Feroz et al. (2008), we can calculate the model selection ratio $R$ then as

\begin{equation}
\label{Eq:muR}
R = \frac{Z_1\mu_{\mathrm{s}}}{Z_0},
\end{equation}
where $\mu_{\mathrm{s}}$ is the expected number of detectable sources in the survey area. This then gives us the probability that a given candidate source is `real' by

\begin{equation}
\label{Eq:prob}
P = \frac{R}{1+R}.
\end{equation}

\section{Source likelihood}
\label{section:likes}

Writing our image as a vector $\mathbf{d}$, we can write the value of a given pixel $d_{i}$ as the sum, 

\begin{equation}
d_{i} = s_{i} + n_{i}.
\end{equation}
where $s_{i}$ is our signal of interest, and $n_{i}$ is an additional noise term.  We model the noise term as a random Gaussian process (although we note that in real datasets non--Gaussianity in the noise can have a significant impact on the statistics, and describe methods to account for this in section \ref{Section:hdf}), and write the probability that our data is described by our model for the signal, which we denote ${\mathbf{s(\Theta)}}$ as,

\begin{equation}
\label{Eq:Prob}
\mathrm{Pr}({\mathbf{d}} | {\mathbf{\Theta}} ) \propto \exp\left[\frac{1}{2}\left({\mathbf{d}} - {\mathbf{s(\Theta)}} \right)^T{\mathbf{C^{-1}}}\left({\mathbf{d}} - {\mathbf{s(\Theta)}} \right)\right]\mathrm{Pr}( {\mathbf{\Theta}})
\end{equation}
where ${\mathbf{\Theta}}$ is the set of parameters that describe our model, and ${\mathbf C}$ is our noise covariance matrix, such that

\begin{equation}
C_{ij}=\langle n_in_j\rangle.
\end{equation}

\subsection{Calculating the Covariance Matrix for Interferometric Images}

We can write our noise vector in the image,

\begin{equation}
{\mathbf{n}}=f^{\dagger}{\mathbf{\tilde{n}}},
\end{equation}
where ${\mathbf{\tilde{n}}}$ is the noise vector in the visibility domain, and $f^{\dagger}$ represents a Fourier transform.  The covariance matrix in the image can then be written

\begin{equation}
{\mathbf{C}} =  f^{\dagger}{\mathbf{\tilde{C}}}f.
\end{equation}
In our likelihood evaluation however we are interested in the inverse of the covariance matrix and so we write 

\begin{equation}
\label{Eq:Eq9}
{\mathbf{C^{-1}}} = f^{\dagger}{\mathbf{\tilde{C}^{-1}}}f.
\end{equation}
In the visibility domain, ${\mathbf{\tilde{C}^{-1}}}$ is diagonal, with elements $\omega_k$ corresponding to the weight of visibility $k$.  Eq. \ref{Eq:Eq9} therefore simplifies to

\begin{equation}
C^{-1}_{ij} =  \sum_{k=1}^{\mathrm{N_{vis}}} \exp{[-i\theta_{ik}]}\omega_k\exp{[i\theta_{jk}]},
\end{equation}
where $\mathrm{N_{vis}}$ is the number of data points in the visibility domain, $\theta_{ik}=2\pi{\mathbf x_i\cdot {\mathbf u_k}}$, with ${\mathbf{u_k}}$ the $(u, v)$ co-ordinates of visibility $k$, and $\mathbf{ x_i}$ the angular separation on the sky of pixel $i$ from the phase centre of the observation.  We are left with our final description of the noise covariance matrix,

\begin{equation}
C_{ij}^{-1} =  \sum_{k=1}^{\mathrm{N_{vis}}} \exp{[-i2\pi ({\mathbf x_j} - {\mathbf x_i})\cdot {\mathbf u_k}]}\omega_k,
\label{Eq:covmatrix}
\end{equation}
which is simply the point spread function (PSF) of the interferometer.

The optimal method for evaluating our likelihood would then be to describe our model $\mathbf{s(\Theta)}$ in the UV domain, and to FFT this using the UV coverage of the observation, calculating the probability of the model according to Eq. \ref{Eq:Prob} and using  Eq. \ref{Eq:covmatrix} as our description of the covariance between the pixels. 
In the following work however we are able to make a set of additional assumptions that allow us to greatly simplify the evaluation of Eq. \ref{Eq:Prob}:

\begin{itemize}

\item Given the resolution of the observations ($\sim$ 3\arcsec or 25kpc at $z=2$) we do not expect to see any extended structure in the image.
\item Given the expected sparsity in the field for sources in our detection regime, we do not expect to have overlapping sources.
\end{itemize}
As such we can take our model in the visibility domain to be a delta function, such that in the image it will be described by the PSF of the interferometer, and we need only be concerned with correlations in the image on the scale of the resolution of the observations.

\subsection{Source Spatial Model}

As we are only concerned with correlations in the image on small scales we model the central region of the PSF as an elliptical Gaussian, with FWHM of the major and minor axes denoted $BMAJ$ and $BMIN$ respectively, and denote the position angle on the sky of the major axis in degrees as $\theta$.  We will justify this assumption in section  \ref{Section:Sim}.  Our model for the source in one channel of the image plane can then be written,

\begin{equation}
s(x,y) = s(v)\exp\left[-\left(a(x - x_0)^2 + 2b(x - x_0)(y - y_0) + c(y - y_0)^2\right)\right],
\end{equation}
with
\begin{eqnarray}
a &=& \frac{\cos^2(\theta)}{2\sigma^2_{\mathrm{maj}}} + \frac{\sin^2(\theta)}{2\sigma^2_{\mathrm{min}}} \\ 
b &=& -\frac{\sin(2\theta)}{4\sigma^2_{\mathrm{maj}}} + \frac{\sin(2\theta)}{4\sigma^2_{\mathrm{min}}} \\
c &=& \frac{\sin^2(\theta)}{2\sigma^2_{\mathrm{maj}}} + \frac{\cos^2(\theta)}{2\sigma^2_{\mathrm{min}}} 
\end{eqnarray}
with $\sigma_{\mathrm{maj}} = BMAJ/2.3548$ and $\sigma_{\mathrm{min}} = BMIN/2.3548$, and $s(v)$ the amplitude of the source in the channel at some velocity $v$.

In order to account for the correlations between the pixels in the image we can define the quantity $I_{\mathrm{sum}}$, where we sum all the pixels for which $s(x,y) > s(v)/2$.  The data can then be described by a single number which we will denote $A_{\mathrm{I}}$, given by
\begin{equation}
\label{Eq:Imsum}
A_{\mathrm{I}} = \frac{\sqrt 2 I_{\mathrm{sum}} }{A_{\mathrm{b}}},
\end{equation}
where $A_{\mathrm{b}}$ is the area of the FWHM of the synthesised beam.  Eq. \ref{Eq:Prob} can then be rewritten for a single channel as,

\begin{equation}
\label{Eq:ImProb}
\mathrm{Pr}({\mathbf{d}} | {\mathbf{\Theta}} ) \propto \exp\left[\frac{1}{2}(A_{\mathrm{I}} - s(v))^2)/\sigma_{\mathrm{I}}^2\right]\mathrm{Pr}( {\mathbf{\Theta}})
\end{equation}
where $\sigma_{\mathrm{I}}$ is the rms of the noise in the image channel. 

\subsection{Source Spectral Model}
\label{Section:SpecModel}

We use three different models for the spectral properties of the source, a Gaussian, double Gaussian, and a more physically motivated model using Brandt's parametrization \citep{1960ApJ...131..293B}.  The first is described simply by a set of three parameters ($v_0$,$\sigma_v$, $A$) as,

\begin{equation}
s(v) = A\exp\left[-\frac{1}{2}(v - v_0)^2/\sigma_v^2\right]
\end{equation}
with $v_0$ the central velocity of the emission line, $\sigma_v$ the scale parameter and $A$ the amplitude at the peak.

The double Gaussian is then described with a set of five parameters ($v_0$,$\sigma_v$, $\Delta v$, $A_1$, $A_2$) as,
\begin{eqnarray}
s(v) &=& A_1\exp\left[-\frac{1}{2}(v - v_0 - \Delta v)^2/\sigma_v^2\right] \\ \nonumber 
     &+&  A_2\exp\left[-\frac{1}{2}(v - v_0 + \Delta v)^2/\sigma_v^2\right]
\end{eqnarray}
where $A_{1,2}$ describe the amplitudes of the two peaks, and $\Delta v$ is the separation of the two components, centered at $v_0$ as before.  Finally, Brandt's parameterisation describes the rotational velocity of a galaxy, modelled as a thin disk,  at some radius $r$ as,

\begin{equation}
V(r) = V_{\mathrm{max}}\frac{\frac{r}{r_{\mathrm{max}}}}{\big[\frac{1}{3} + \frac{2}{3}\big(\frac{r}{r_{\mathrm{max}}}\big)^n\big]^{\frac{3}{2n}}},
\end{equation}
where $r_{\mathrm{max}}$ is the radius at which the galaxy obtains it's maximum rotational velocity $V_{\mathrm{max}}$, and $n$ describes how quickly the rotational velocity drops away with radius.

The line profile can then be calculated following the same method as in \citet{2009ApJ...698.1467O}; Starting with a thin ring, the observed circular velocity at any point will be given by $V_{\mathrm{c}}\sin\gamma$, so that the luminosity density $\phi$ as a function of $V_{\mathrm{obs}}$ will be given by:

\begin{equation}
\phi (V_{\mathrm{obs}},V_{\mathrm{c}}) = \frac{d\gamma}{dV_{\mathrm{obs}}} = \frac{1}{\sqrt{V_c^2 - V_{\mathrm{obs}}^2}} \;\;\;\;\mathsf{for}\; |V_{\mathrm{obs}}| < V_{\mathrm{c}}.
\end{equation}
This however is divergent as $V_{\mathrm{obs}}$ approaches $V_\mathrm{c}$, and so we need to smooth it by taking into account the normal distribution of velocities produced by the random motions in the gas.  In Eq. \ref{Eq:gasdisp} the gas dispersion $\sigma_{\mathrm{gas}}$ is taken to be constant with a value of $8~\mathrm{km~s^{-1}}$, however the exact value is not critical, so that the final velocity profile is given by:

\begin{equation}
\label{Eq:gasdisp}
\Phi(V_{\mathrm{obs}}, V_{\mathrm{c}}) = \frac{1}{\sqrt{2\pi}\sigma_{\mathrm{gas}}}\int\;\mathrm{d}V\exp\bigg[\frac{(V_{\mathrm{obs}}-V)^2}{-2\sigma_{\mathrm{gas}}^2}\bigg]\phi(V,V_{\mathrm{c}})
\end{equation}
Assuming an exponential profile for the gas within the disk, described by some scale radius $r_{\mathrm{scale}}$ the emission line $\Phi (V_{\mathrm{obs}})$ can then be modelled by integrating:

\begin{equation}
\Psi (V_{\mathrm{obs}}) = \int_0^{\infty}\;r\exp(-r/r_{\mathrm{scale}})\Phi(V_{\mathrm{obs}},V_{\mathrm{c}}(r)) \mathrm{d}r
\end{equation}
This function is then normalized to have a peak of unity to facilitate later scaling by the amplitude, such that

\begin{equation}
s(v)  = A\Psi (v).
\end{equation}

\subsection{The final likelihood}

Given a spectral model $s(v)$ we can then write the final likelihood that we will use in our search as a product over a set of channels where the likelihood in any one channel is given by Eq \ref{Eq:ImProb}.  Our final probability is therefore given by:

\begin{equation}
\label{Eq:ImProbFinal}
\mathrm{Pr}({\mathbf{d}} | {\mathbf{\Theta}} ) \propto \mathrm{Pr}( {\mathbf{\Theta}}) \prod_{i=1}^{\mathrm{N_{chan}}} \exp\left[\frac{1}{2}(A_i - s(v_i))^2)/\sigma_i^2\right],
\end{equation}
where $N_{\mathrm{chan}}$ is the number of channels in the image.

\subsection{Comparison to the SERCH algorithm}

We can briefly consider the two main differences between Eq. \ref{Eq:ImProbFinal} and the SERCH algorithm.  SERCH performs a weighted coherent sum over the the channels in the image, where the weighting is defined by a gaussian of some width.

Denoting the normalised model gaussian in velocity space such that the peak of the model is equal to 1 as $\bar{s}(v)$, we can write this sum for any pixel in the image $d_i$ as:

\begin{equation}
\label{Eq:dhat}
\hat{d}_i = \sum_{j=1}^{\mathrm{N_{chan}}}d_i(v_j)\bar{s}(v_j),
\end{equation}
where $v_j$ is the reference velocity for channel $j$.  Similarily the noise in the summed image will be

\begin{equation}
\label{Eq:nhat}
\hat{\sigma} = \sum_{j=1}^{\mathrm{N_{chan}}}\sigma_j\bar{s}(v_j),
\end{equation}
and the expected signal will simply be
\begin{equation}
\hat{s} = \sum_{j=1}^{\mathrm{N_{chan}}}s(v_j).
\end{equation}
Using this notation we could then rewrite Eq. \ref{Eq:ImProb} as:

\begin{equation}
\label{Eq:ImProbSerch}
\mathrm{Pr}({\mathbf{d}} | {\mathbf{\Theta}} ) \propto \exp\left[\frac{1}{2}(\hat{A} - \hat{s})^2)/\hat{\sigma}^2\right]\mathrm{Pr}( {\mathbf{\Theta}}),
\end{equation}
where $\hat{A}$ is calculated as in Eq. \ref{Eq:Imsum}, but now in the summed image. This has the disadvantage however that it does not correctly take into account large positive spikes in the noise, as it will simply add to the S:N of the final summed image.   In Eq. \ref{Eq:ImProbFinal} this cannot happen, as the difference between the model and the noise spike will be large if the surrounding channels do not also support a model with high amplitude, and so the candidate will be appropriately down weighted.  Secondly Eq \ref{Eq:Imsum} takes better account for the correlation between pixels in the image, averaging over the values within the FWHM of the PSF, and will therefore be more robust to individual outliers that will trigger a detection if the S:N threshold is set on a per pixel basis.

\section{Application to a Simulated Survey}
\label{Section:Sim}

We now apply our search technique to a simulated data set, in order to test both the robustness of parameter estimation, and to determine a suitable probability threshold $p_{\mathrm{th}}$ for use in the analysis of the survey data such that, if the probability given by Eq. \ref{Eq:prob} is greater than $p_{\mathrm{th}}$, we consider it to be a real source.

\subsection{The Simulated Data Set}

In order to make the simulation as applicable to our survey as possible, we use the UV sampling function from the observation described in section \ref{Section:hdf} to create a set of empty UV data points.  In order to recreate the same variation in noise across baselines, frequency and time present in the survey we add uncorrelated Gaussian noise to each UV point with rms determined by the weight of that data point in the survey data.

A set of 100 model galaxies are then added to the simulation, uniformly distributed in both space and frequency. These sources are described spatially by a circular Gaussian, and use the brandt parameterization in velocity space, as described in section \ref{Section:SpecModel}.  With the exception of source amplitude the model parameters are chosen uniformly across the priors described in Table \ref{table:priors}.   Source amplitudes are then chosen such that the distribution of model flux follows an arbitrary power law.  Setting a detection limit such that the integrated signal to noise of a given source within the data is greater than 4, this results in a value of $\mu_{\mathrm{s}}$ of 22.

\begin{table}
\centering
\caption{Parameter ranges used for the simulated galaxies.} 
\centering 
\begin{tabular}{c c c c} 
\hline\hline 
Prior & Unit & Min & Max\\ [0.5ex] 
\hline 
$\sigma_{1,2}$ &arcsec & 0.1 & 1.0 \\
$V_{\mathrm{max}}$ & km~$\mathrm{s^{-1}}$ & 40& 500 \\
$\sigma_\mathrm{v}$ & km~$\mathrm{s^{-1}}$ & 40 & 400 \\
$r_{\mathrm{max}}$ & kpc & 0.1 & 10.0 \\
n & none & 0.1 & 3.0 \\
$r_{\mathrm{scale}}$ & kpc & 0.1 & 10.0 \\ 
\hline 
\end{tabular}
\label{table:priors} 
\end{table}

These 22 sources span a range of physical parameters, with velocity widths ranging from 1 to 10 channels, and peak amplitude per channel dropping to as low as twice the noise level in that channel.  

Finally in order to reproduce some of the usual effects that must be dealt with in interferometric data, gain errors are introduced on 2 random baselines, such that the values of the data points on those baselines are increased by 20$\%$, a source of error not modeled in the following analysis.

This data set is then used to generate an image using natural weighting, with pixel size 0.5$\arcsec$ in which the search is performed.  We also fourier transform the UV sampling function itself in order to test our assumption that the PSF is well described by a Gaussian.   Table \ref{Table:PSF} lists the properties of the Gaussian approxmation to the point spread function (PSF) for each of the spectral windows included in the simulation  In each case the recovered amplitude is calculated using Eq. \ref{Eq:Imsum}.  We find that the maximum deviation from the expected value of 1 is $\sim 2-3\%$, and therefore consider the approximation valid for the dataset. 
Finally, in performing the search we use only the Gaussian or double Gaussian spectral models in order to eliminate the introduction of any bias from using a matched filter in the search.

\begin{table}
\centering
\caption{Properties of the PSF for different spectral windows} 
\centering 
\begin{tabular}{c c c c c c} 
\hline\hline 
spectral window & BMAJ & BMIN & BPA & Recovered Amplitude\\ [0.5ex] 
	GHz 	&  arcsec	&arcsec  &	 degrees  &  arbitrary units\\
\hline 
81.5	& 3.47 &  3.04 &  73.71 & 1.02\\
83.3	& 3.48 &  3.03 &  72.93 & 1.01 \\
85.0	& 3.74 &  3.23 &  69.54 & 1.03\\
86.9	& 3.74 &  3.22 &  69.31 & 1.02\\
88.5	& 3.47 &  2.83 &  88.35 & 1.01\\
90.5	& 3.50 &  2.80 &  87.57 & 1.00\\
92.0	& 3.35 &  2.88 &  92.25 & 1.02\\
94.0	& 3.35 &  2.86 &  89.83 & 1.02\\
95.5	& 3.25 &  2.73 &  82.07 & 1.02\\
97.6	& 3.26 &  2.70 &  80.46 & 1.01\\
99.0	& 3.49 &  2.72 & 101.66 & 1.03\\
101.2	& 3.54 &  2.73 &  99.84 & 1.02\\
102.5	& 2.93 &  2.55 &  67.75 & 1.02\\
104.8	& 2.95 &  2.56 &  66.85 & 1.01\\
106.0	& 2.83 &  2.46 &  69.63 & 1.01\\
107.9	& 2.82 &  2.47 &  67.72 & 1.01\\
109.5	& 3.24 &  2.60 &  63.30 & 1.02\\
111.4	& 3.26 &  2.58 &  63.43 & 1.02\\
113.0	& 2.90 &  2.60 &  86.64 & 1.00\\
115.0	& 2.88 &  2.60 &  84.52 & 1.00\\
\hline 
\end{tabular}
\label{Table:PSF} 
\end{table}

\subsubsection{Completeness and Purity}

\begin{figure*}
\begin{minipage}{168mm}
\begin{center}$
\begin{array}{cc}
\includegraphics[width=80mm]{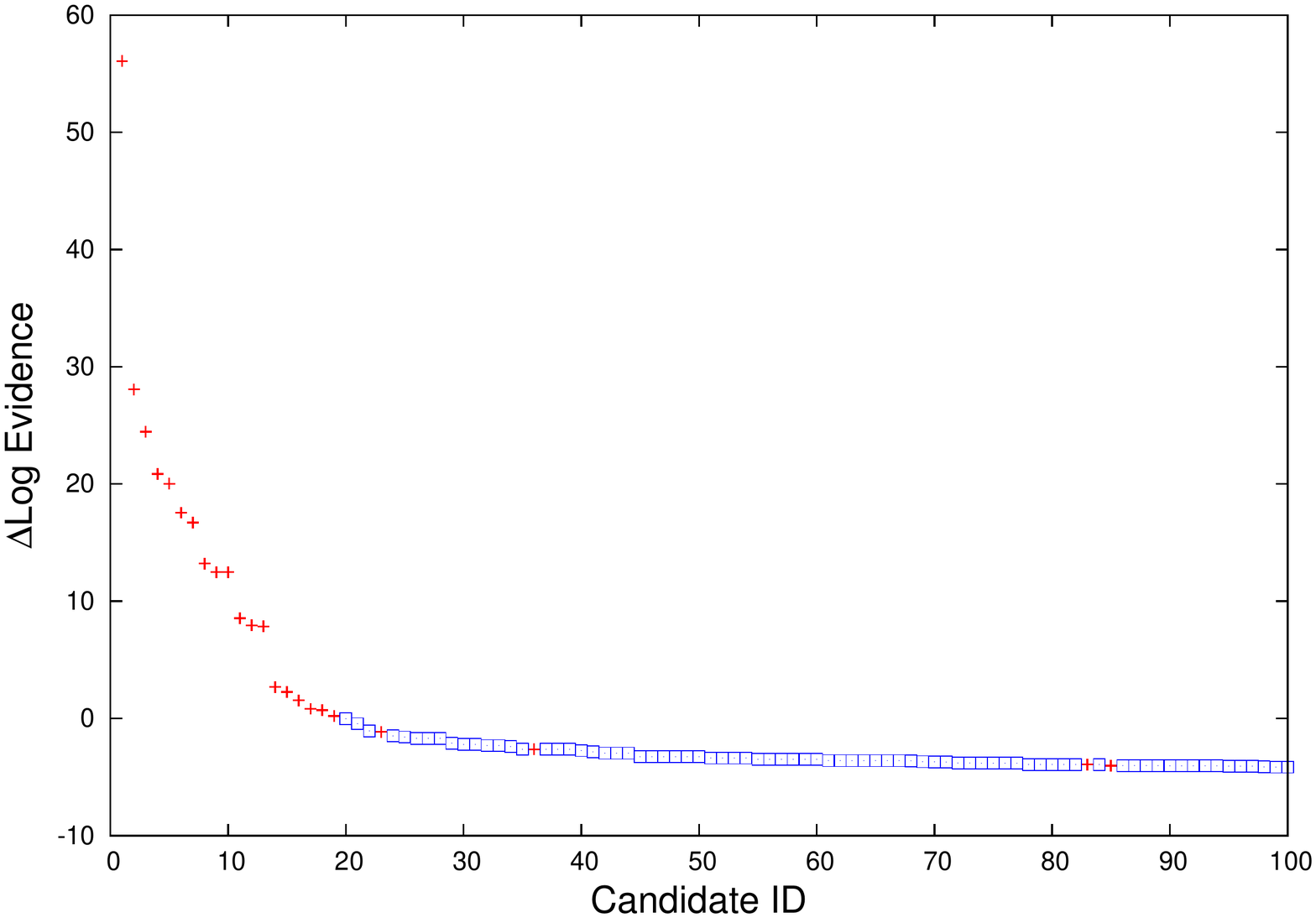} &
\includegraphics[width=80mm]{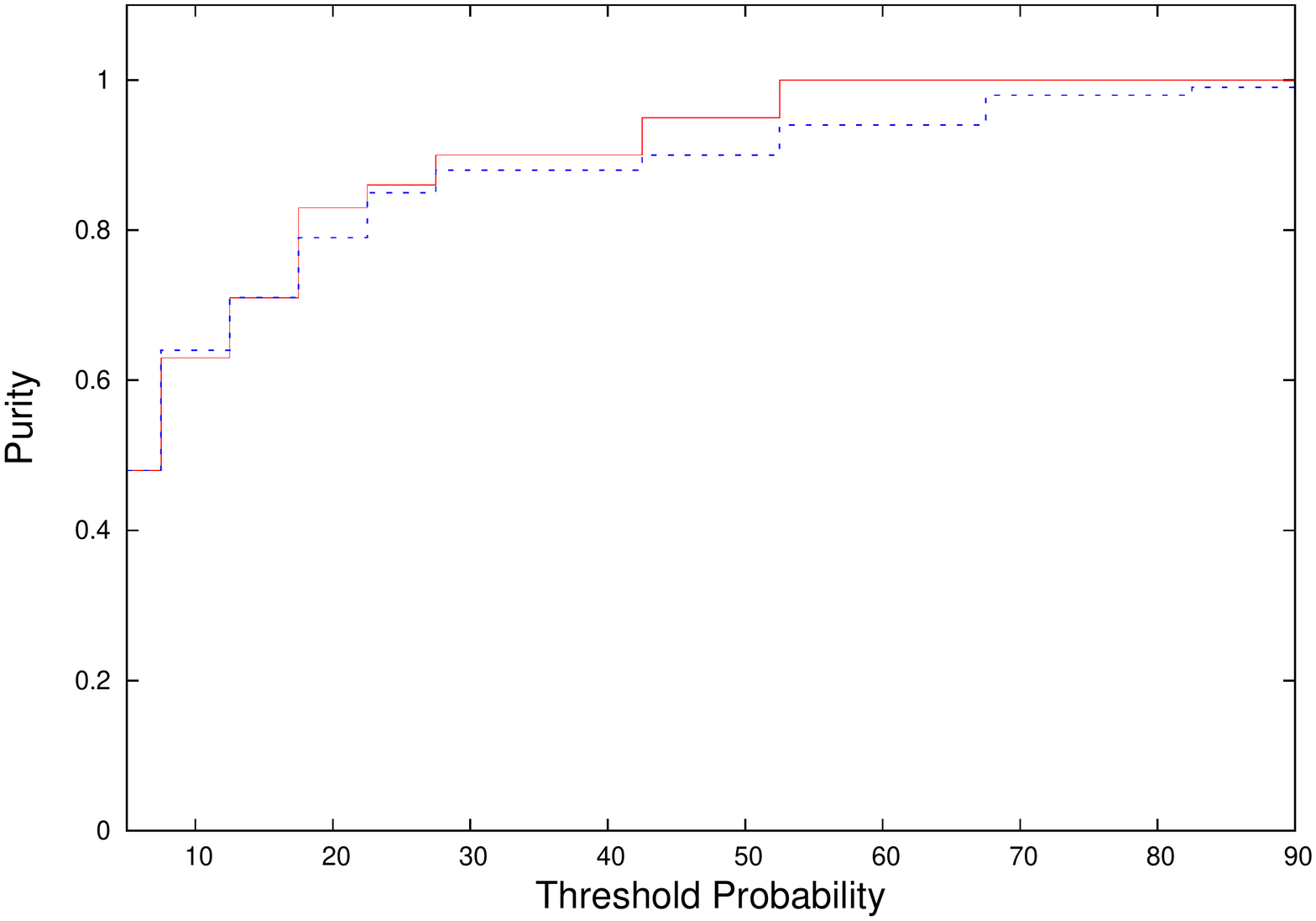} 
\end{array}$
\end{center}
\caption{Left: $\Delta \log $ Evidence for the 100 most probable source candidates comparing the evidence for a source being present and for no source being present. red crosses indicate the source corresponds to an entry in the catalogue, blue square points indicate that the source does not. The blue line indicates a probability of 50$\%$. Right: Actual (Red solid) and theoretical (Blue dashed) values for the purity of our source list as a function of probability threshold.}
\label{Fig:Purity}
\end{minipage}
\end{figure*}

MULTINEST found a total of 2387 and 1951 candidate objects in the image for the Gaussian and double Gaussian models respectively.  These candidates were then cross matched with entries in the simulated catalogue to see if they corresponded to a true source.  Matching was performed by identifying those candidates with central frequencies and positions that were within 2 channels and 3 arcseconds of the catalogue values.  In cases where multiple matches were made to a single catalogue source, the model with the greatest Evidence was taken to represent the source. The left plot in figure \ref{Fig:Purity} shows the probability of the top 100 candidates returned by MULTINEST ordered by decreasing evidence.   A green point indicates a match was identified from the catalogue, and red that no match was found.  In order to determine a suitable threshold probability we consider the purity of the returned samples as a function of the threshold used.  The purity is defined as the ratio of detections that correspond to a true source over the total number of detections above the threshold $p_{\mathrm{th}}$.  A comparison between the actual purity (in red), and the theoretical purity (in green) is shown in the right plot of figure \ref{Fig:Purity}.  We calculate the theoretical purity by considering the total number of expected false detections $N_{\mathrm{f}}$ within our results for a given $p_{\mathrm{th}}$ given by the following sum:

\begin{equation}
\label{Eq:FalseAlarm}
N_{\mathrm{f}} = \sum_{i=1}^{N_{\mathrm{tot}}} 1 - P_i
\end{equation}
where $N_{\mathrm{tot}}$ is the total number of sources above our acceptance threshold, and $P_i$ is the probability of each.

For a threshold probability of 50$\%$ we find an actual purity of 95$\%$.  Below this point the purity begins to decline slowly, reaching $\sim$ 85 $\%$ with $p_{\mathrm{th}} = 25\%$ and falling off rapidly thereafter.  Both the actual and theoretical purities are consistent with one another across the full range of $p_{\mathrm{th}}$.

Of the 22 sources that have theoretical integrated signal to noise of greater than 4, 18 are detected with $p_{\mathrm{th}} = 50 \%$ giving a completeness ratio of 82$\%$.  Each of the 4 undetected sources has an actual signal to noise ratio of less than 4 in the simulated data, with the greatest being 3.84.  Recovered galaxies within the data exhibit a range of line widths between 1 and 10 channels, with signal to noise per channel ranging between 2 and 6, representing a complete recovery of what can realistically be expected in the actual survey data.   Decreasing $p_{\mathrm{th}}$ to 20$\%$ the completeness increases only marginally to 86$\%$ and so for the purposes of this study we set $p_{\mathrm{th}}$ to be 50$\%$ in order to maintain an acceptable balance between purity and completeness.

\subsubsection{Parameter Estimation}

\begin{figure*}
\begin{center}$
\begin{array}{c}
\vspace{-1.9cm}
\includegraphics[width=100mm]{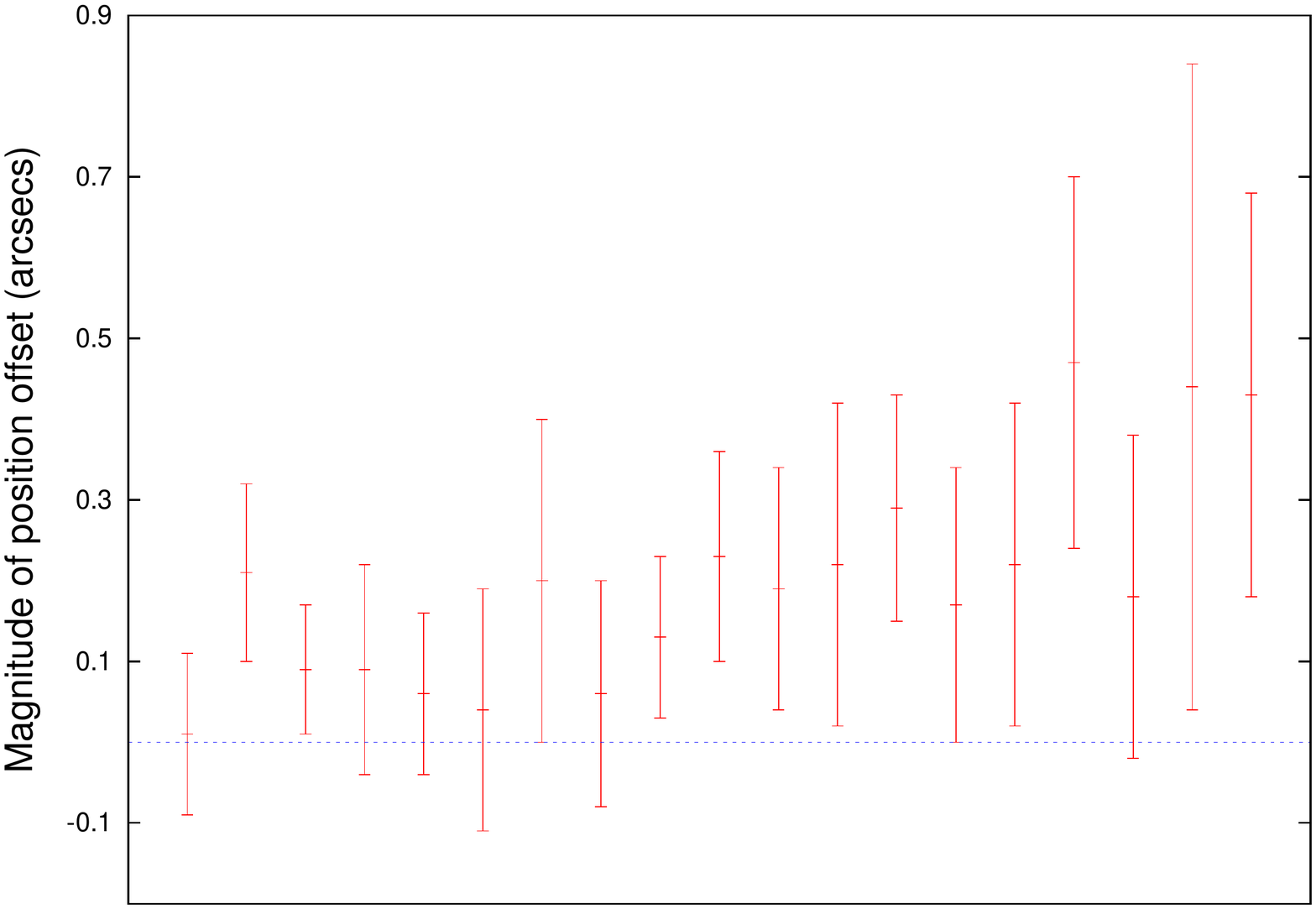} \\
\vspace{-1.88cm}
\includegraphics[width=100mm]{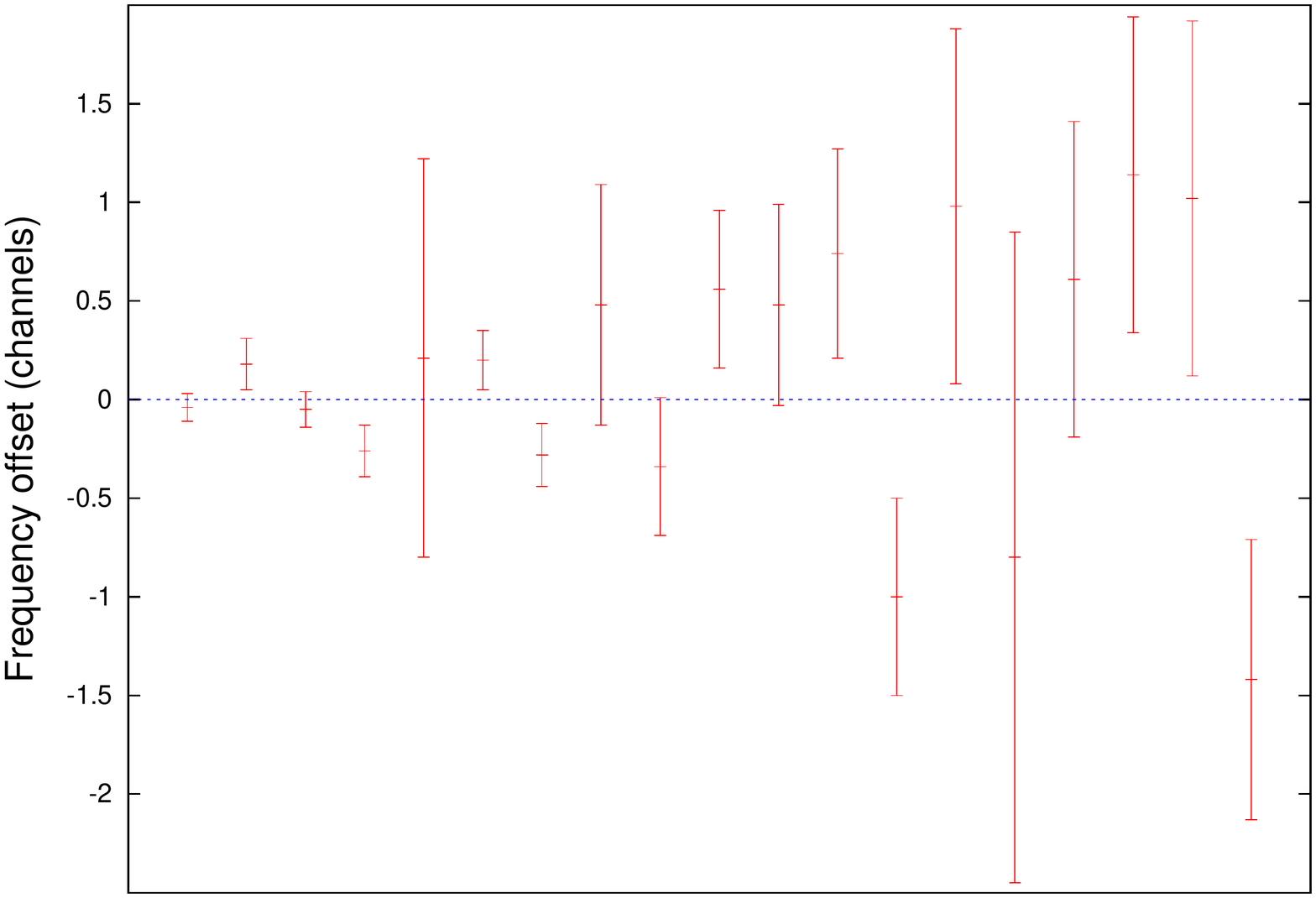} \\
\includegraphics[width=100mm]{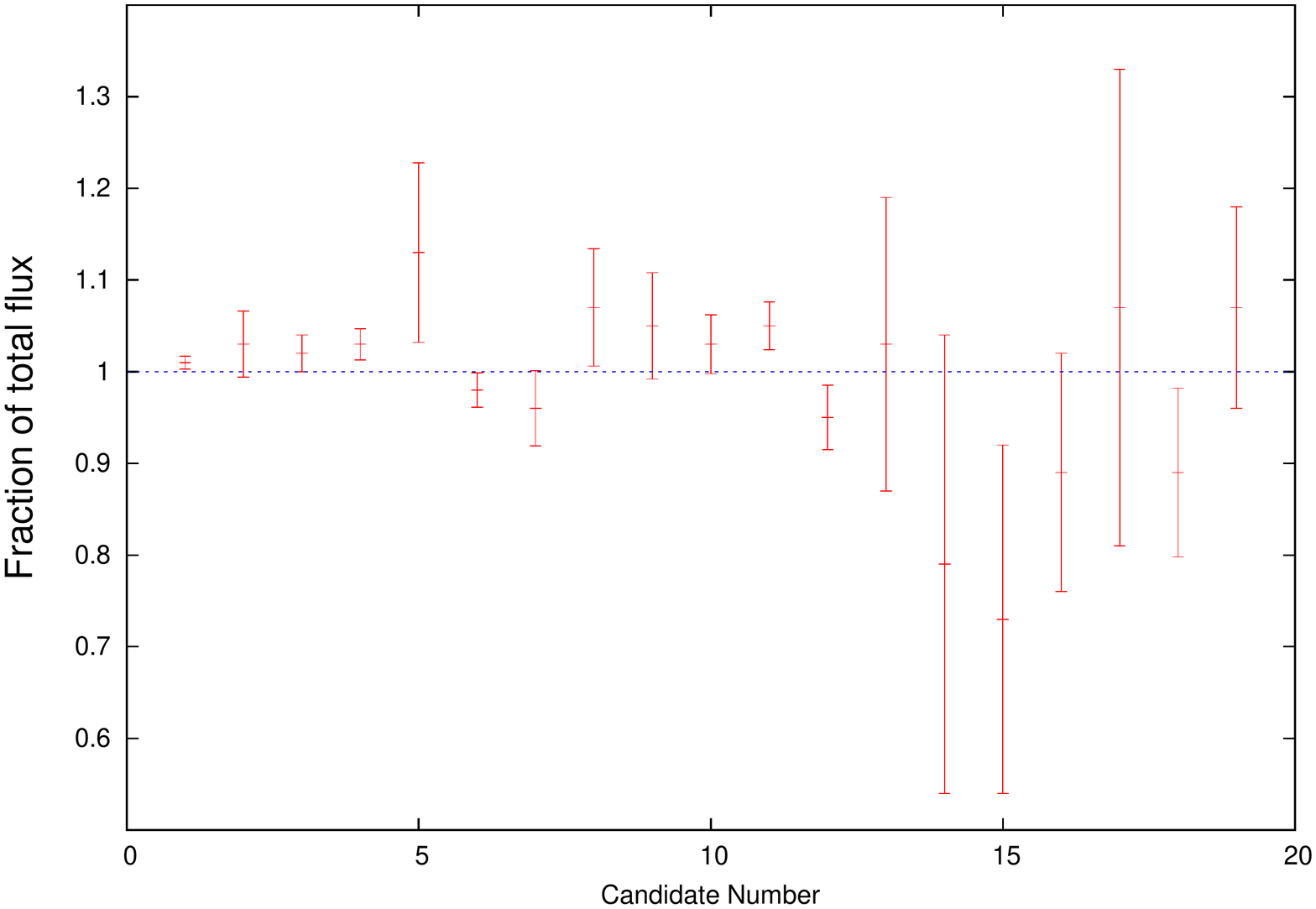} \\
\end{array}$
\end{center}
\caption{From top to bottom, parameter offsets and 1$\sigma$ uncertainties for spatial position, frequency and fraction of total flux for the 19 source candidates returned by the simulated blind search.}
\label{figure:seppar}
\end{figure*}

We now compare a subset of the estimated parameters for those galaxy models with probabilities greater than $p_{\mathrm{th}}$ that have a confirmed counterpart in the source catalogue.  In Fig \ref{figure:seppar} we show the magnitude of the separation in spatial position in arcseconds (top), central frequency in channels (middle) and total model flux in terms of the fraction of the total model flux (bottom) between the mean parameter estimate for each source and the injected values.  In each case the error bars represent the one sigma errors returned by MULTINEST during the search.

In all cases the parameter estimates are consistent to within $\sim 2\sigma$ errors, with frequency and spatial positions recovered with an average difference of 0.22 arcseconds ($\sim$ half a pixel width) and 0.55 channels respectively between the model and the injected values.  We note that the uncertainties on the spatial positions increase steadily with S:N as would be expected given the size of the sources simulated is always much smaller than the PSF in the image domain they will all have the same structure.  In comparison the uncertainties on total flux and central frequency show more variation, but the variation is correlated between the two parameters, i.e. the wider sources that take up more channels have greater uncertainty in their central frequency, however they also have the greatest uncertainty in their total flux as the total integrated noise will also be greater across the increased number of channels.

\section{The Hubble Deep Field Survey}
\label{Section:hdf}

We now apply our Bayesian search technique to a molecular line scan in the Hubble Deep Field North.  Details of the observation are given in D14, however we repeat the key details here to aid future discussion.  The scan covers the entire 3mm window (79-115~GHz) using the IRAM PdBI with approximately uniform sensitivity, with average noise of $\sim 0.3~\mathrm{Jy beam^{-1}}$ in a $90~\mathrm{km s^{-1}}$ channel.   The primary beam of the PdBI can be described by a Gaussian profile with $\mathrm{FWHM} = 47.3" \times (100/\nu)$, where $\nu$ is the observing frequency in GHz.  Observations were performed in C-array configuration, with average beam size of $\sim 3\arcsec$, corresponding to $\sim 20$ kpc at redshifts $\ge 1$, as such we do not expect to spatially resolve high-redshift galaxies.

We perform our search within a radius of $\sqrt2$ times the radius of the primary beam.  In total the observation covers a cosmic volume of $\sim$ 7000 $\mathrm{Mpc^3}$ for redshift ranges of $z < 0.45, 1.01 <z<1.89, z>2$ for different rotational transitions of CO.

\subsection{Expected Source Counts}
\label{Section:Expected}

As discussed in Section \ref{Section:mu} in order to assign a final probability to source candidates we require an estimate of the expected number of detections within the field above some threshold with which to normalise those probabilities.  This acts to quantify the notion that we are more likely to believe a less significant detection if the search volume around that source is smaller because the probability of spurious source-like noise decreases as the search volume per source decreases.

In D14 the expected source counts in the observation given the sensitivity quoted is discussed in relation to existing theoretical estimates, in particular the expected distributions of low-J CO transitions based on predictions by \cite{2013arXiv1303.4392S,2012ApJ...747L..31S} and \cite{2013ApJ...765....9D}.  These predictions suggest that we should expect only $\sim 1-2$ detections within the data.  Including the known CO($5\to 4$) and CO($6\to 5$) emission lines associated with the SMG HDF850.1 from these calculations the expected number of detections $\mu_{\mathrm{s}}$ in the data will be $\sim 3-4$.  Ultimately however the purpose of this survey was to determine this density observationally in an unbiased way.  As such assuming one specific value for $\mu_{\mathrm{s}}$ in our results might lead us to discount source candidates that otherwise would be included in the final catalogue.  We therefore calculate final probabilities for a range of $\mu_{\mathrm{s}}$, allowing values from $\mu_{\mathrm{s}}=3$ to $\mu_{\mathrm{s}}=7$.  In order to determine an appropriate final value we then randomise the order of the channels twice and repeat the analysis each time.  By seting our prior on the width of the source candidates to be a minimum of two channels ($\sim 180 ~\mathrm{km s^{-1}}$) any sources detected in these shuffled datasets will therefore be spurious noise detections that can be used to confirm our Bayesian false alarm rate.

\subsection{The Blind Survey}
\label{Section:Blind}

\begin{figure*}
\begin{center}$
\begin{array}{cc}
\hspace{-1.5cm}
\includegraphics[width=100mm]{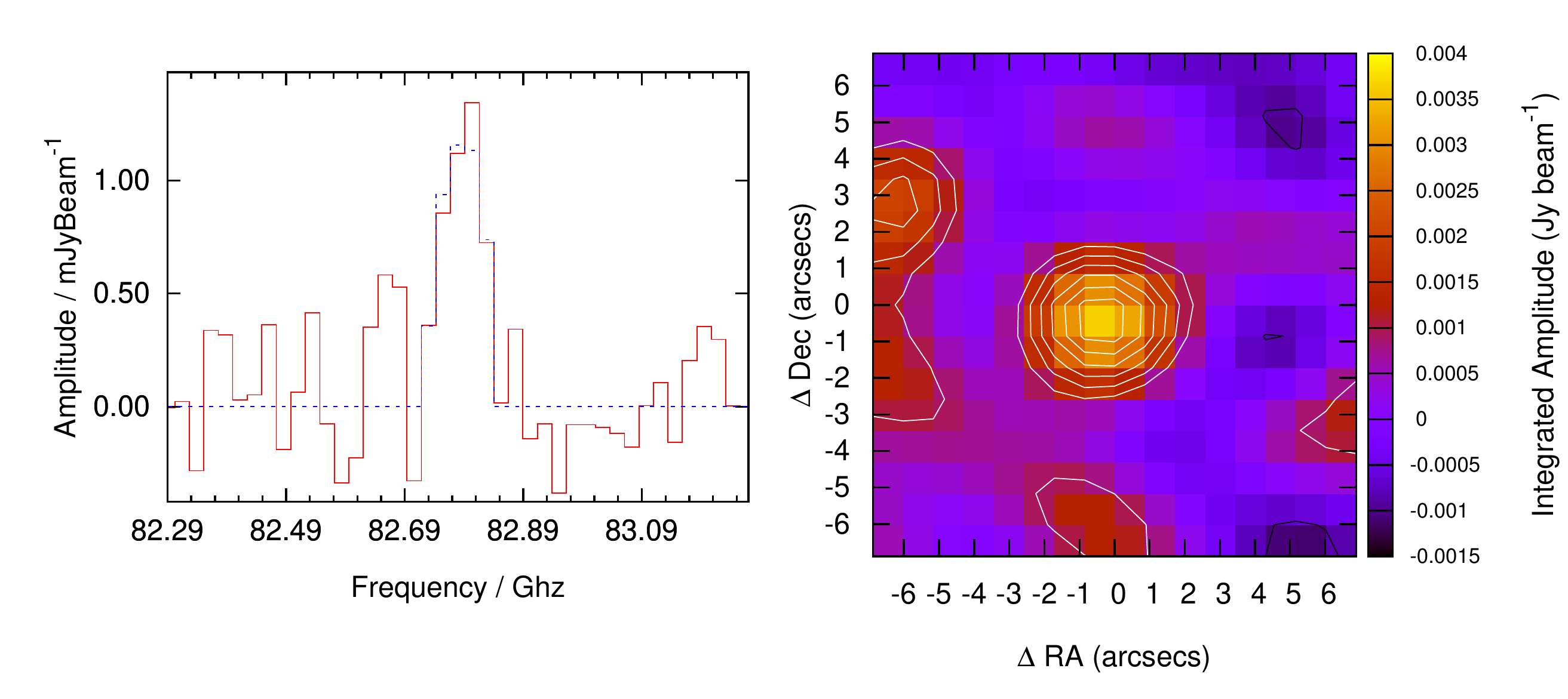} &
\includegraphics[width=100mm]{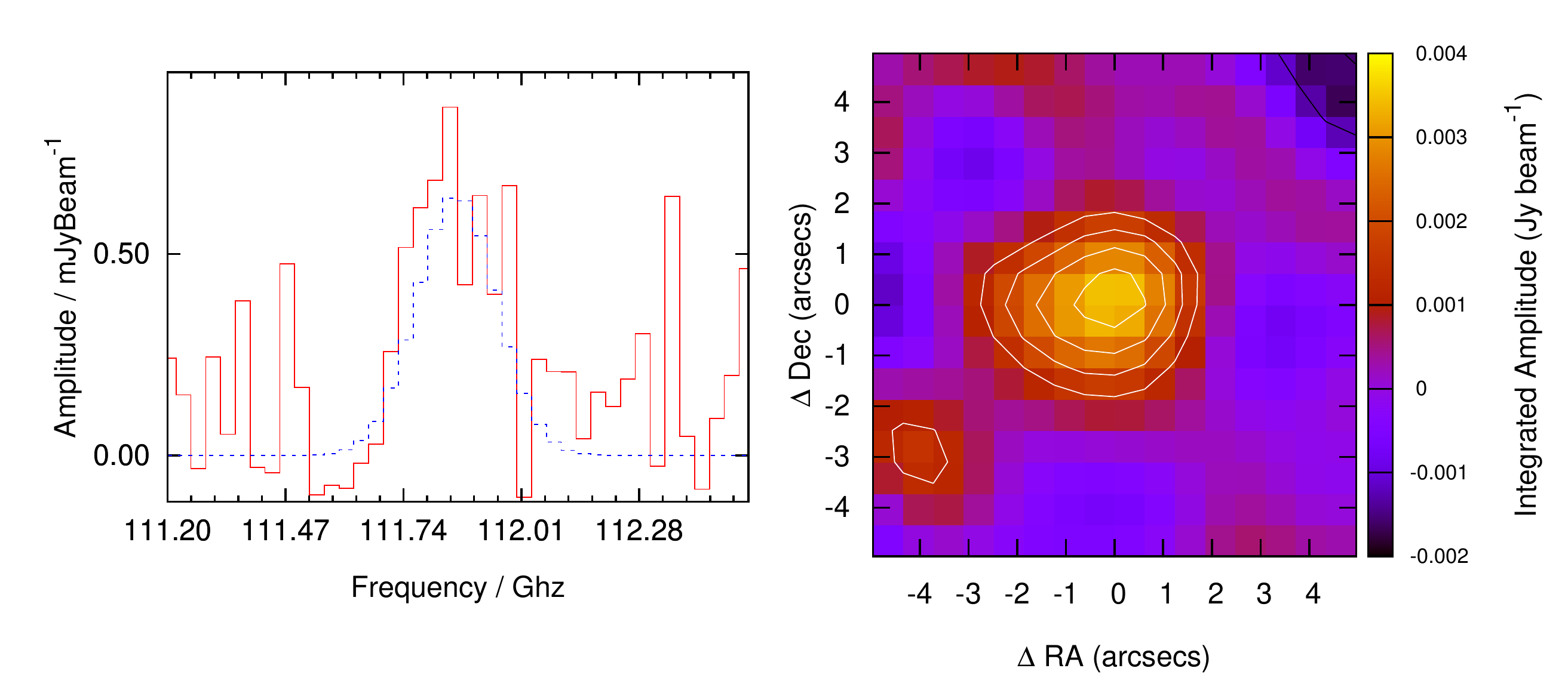} \\
\hspace{-1.5cm}
\includegraphics[width=100mm]{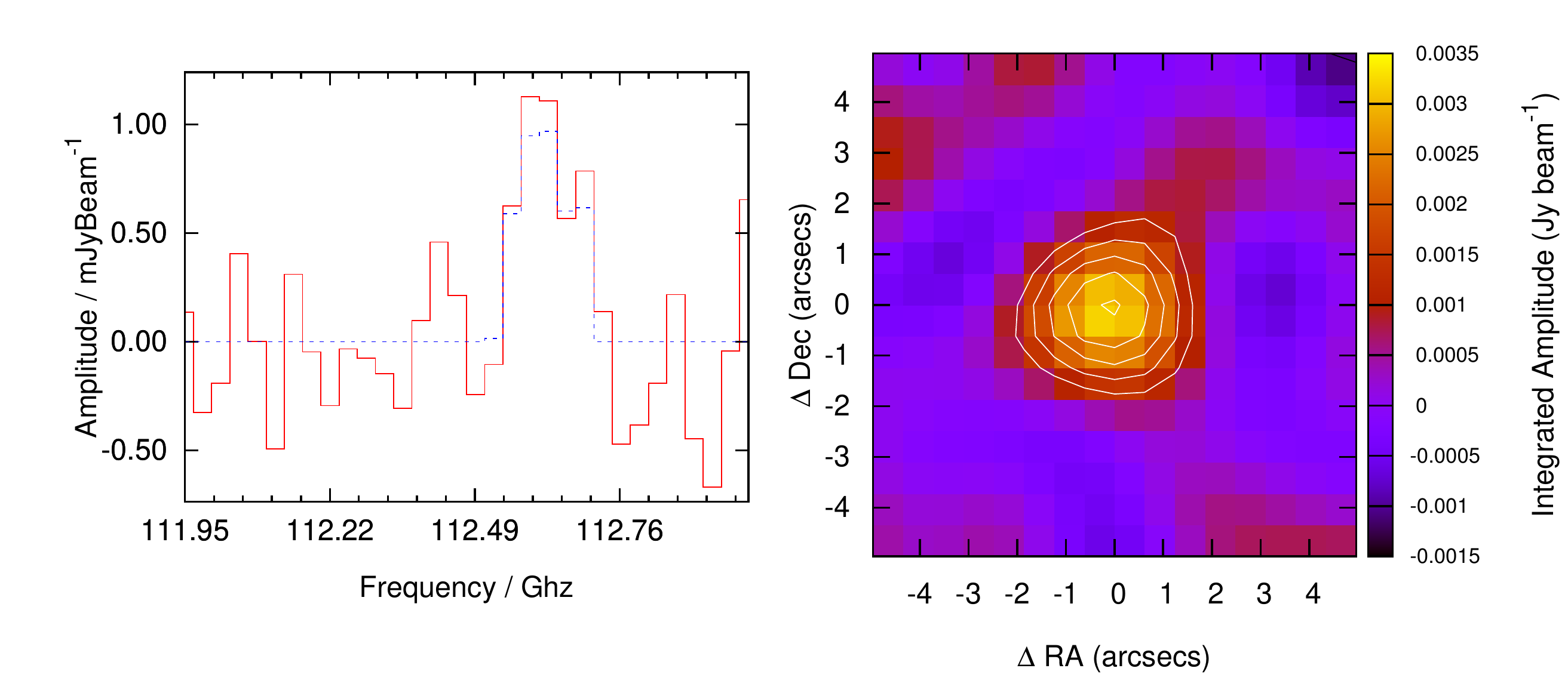} &
\includegraphics[width=100mm]{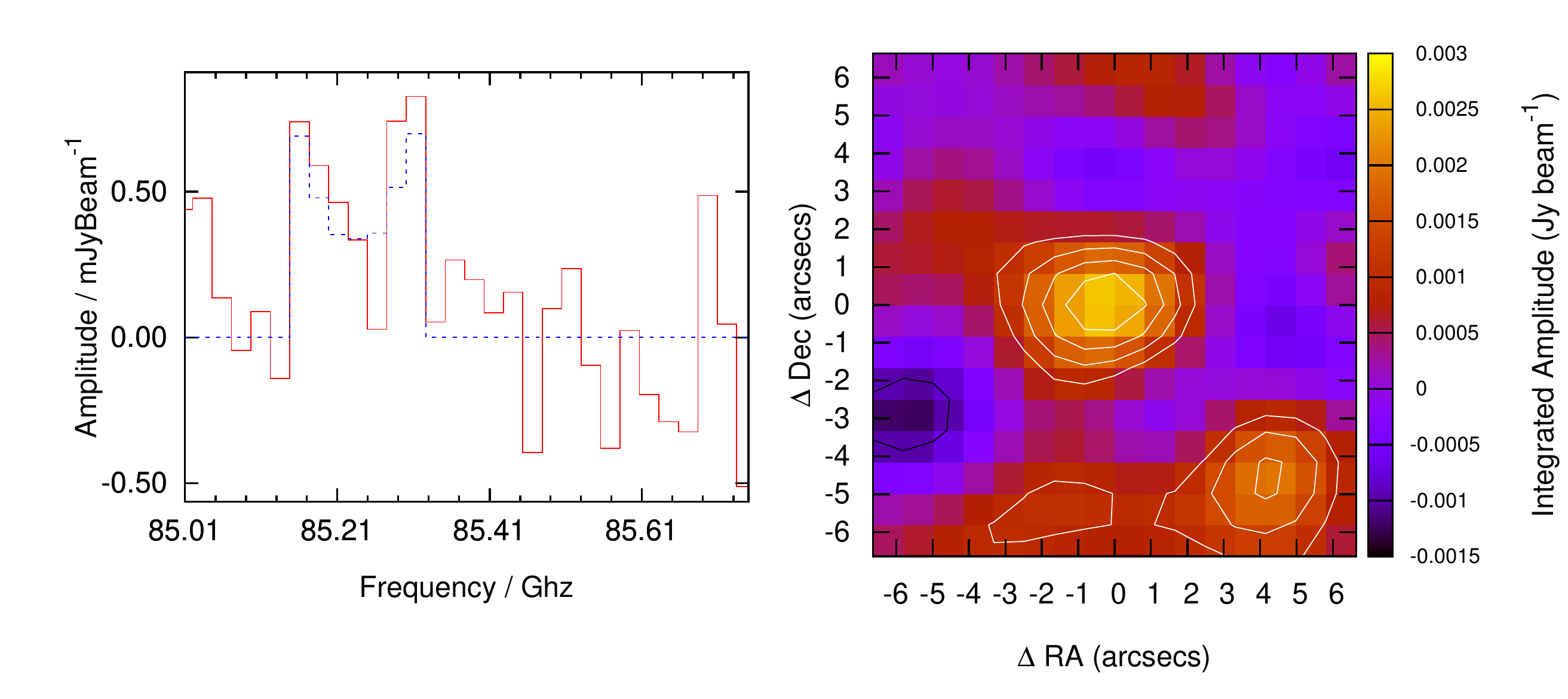}\\ 
\hspace{-1.5cm}
\includegraphics[width=100mm]{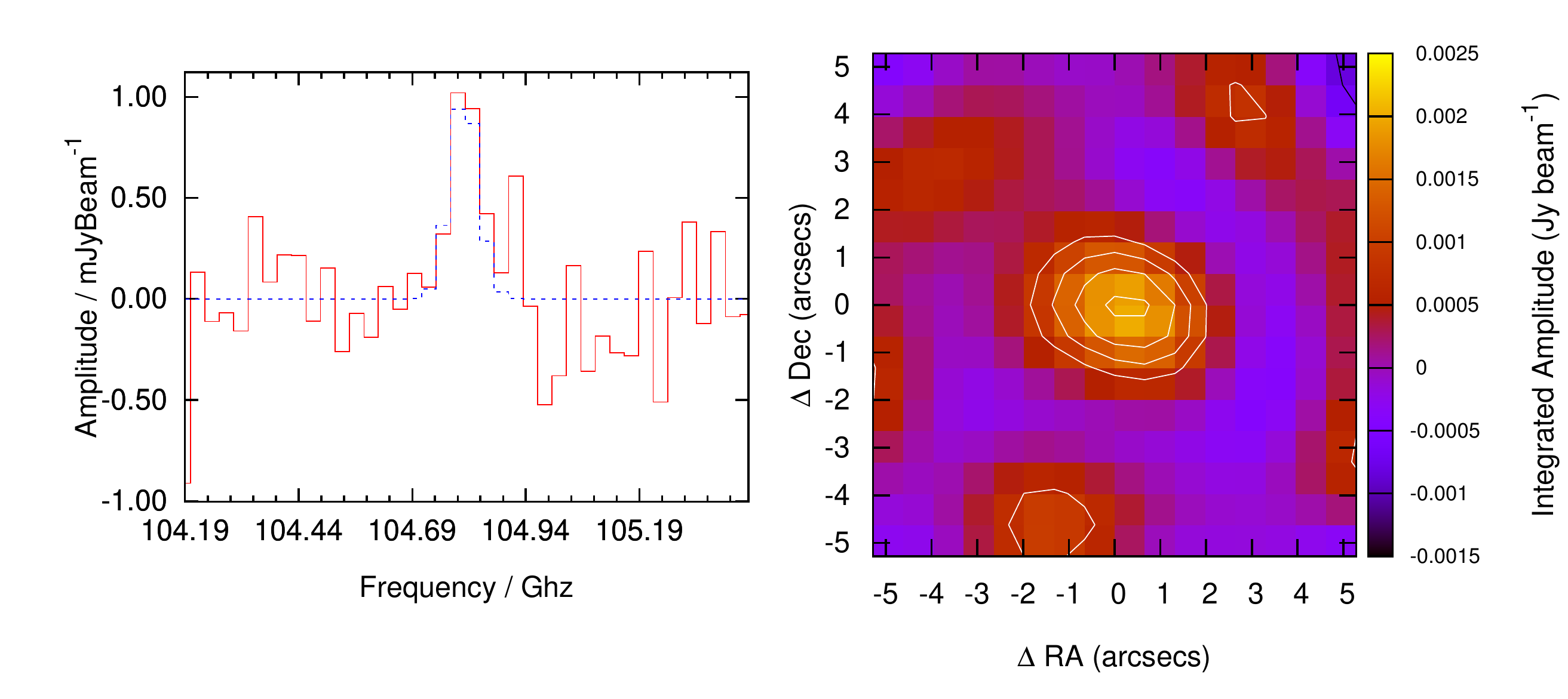} &
\includegraphics[width=100mm]{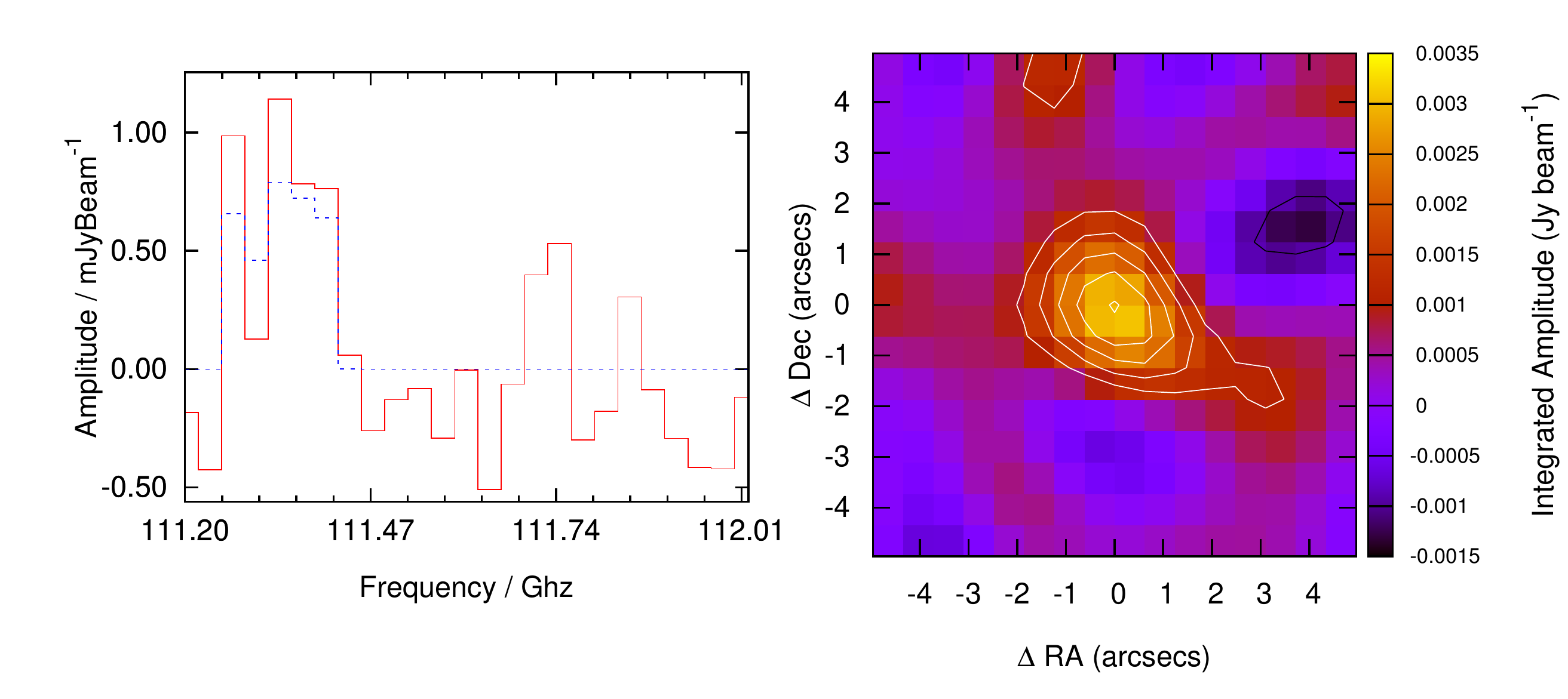}
\end{array}$
\end{center}
\caption{From left to right, top to bottom, maps and spectra for candidate sources B1-6 from Table \ref{Table:MainResults}. Spectra and maps are centered at the nominal location of the candidate given in Table \ref{Table:MainResults}, unless they are close to the edge of a spectral window in which case they will appear as slightly offset.  Spectra show the data (red solid line) with the best fit spectral model overlaid (blue dotted line).  Maps are produced by integrating over those channels in which the model amplitude of the source is greater than twice the noise level in the image, with units in Jy~beam$^{-1}$.   Postive contours are shown in white, negative contours in black, with initial contours at $\pm$2$\sigma$ increasing by 1 each time.}
\label{figure:blind}
\end{figure*}

Table \ref{Table:MainResults} summarizes some of the parameters for the candidate galaxies detected with probabilities greater than our threshold for acceptance of 50$\%$ when using different values for the expected number of sources in the field $\mu_{\mathrm{s}}$ denoted B$X$.  Where these candidates correspond to a source ID in D14 we also list the corresponding identifier [ID$X$].  For each of these sources we list the best fit positions and frequencies, FWHM, the primary beam corrected velocity integrated CO flux ($I_{\mathrm{CO}}$), signal to noise ratio of the source defined in terms of Eq. \ref{Eq:ImProbFinal} such that S:N=$\hat{A} /\hat{\sigma}$, where the kernal used for the weighted sum is the mean source model returned for each candidate, and finally its probability defined by Eq. \ref{Eq:prob}. Spectra and maps are shown for each of the sources where $\mu_{\mathrm{s}}=3$ in Fig. \ref{figure:blind}. 

\begin{figure*}
\begin{center}$
\begin{array}{cc}
\hspace{-1.5cm}
\includegraphics[width=100mm]{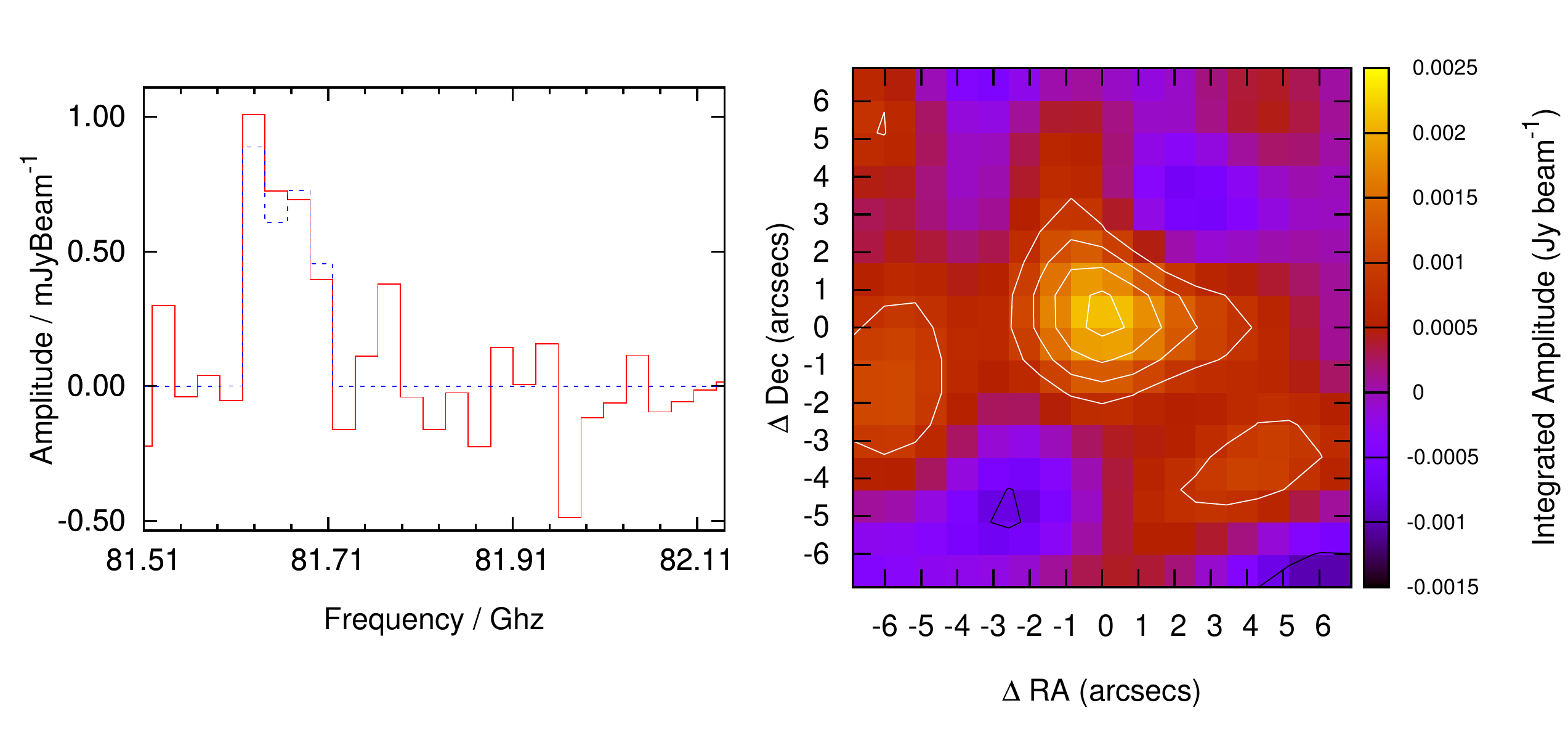} &
\includegraphics[width=100mm]{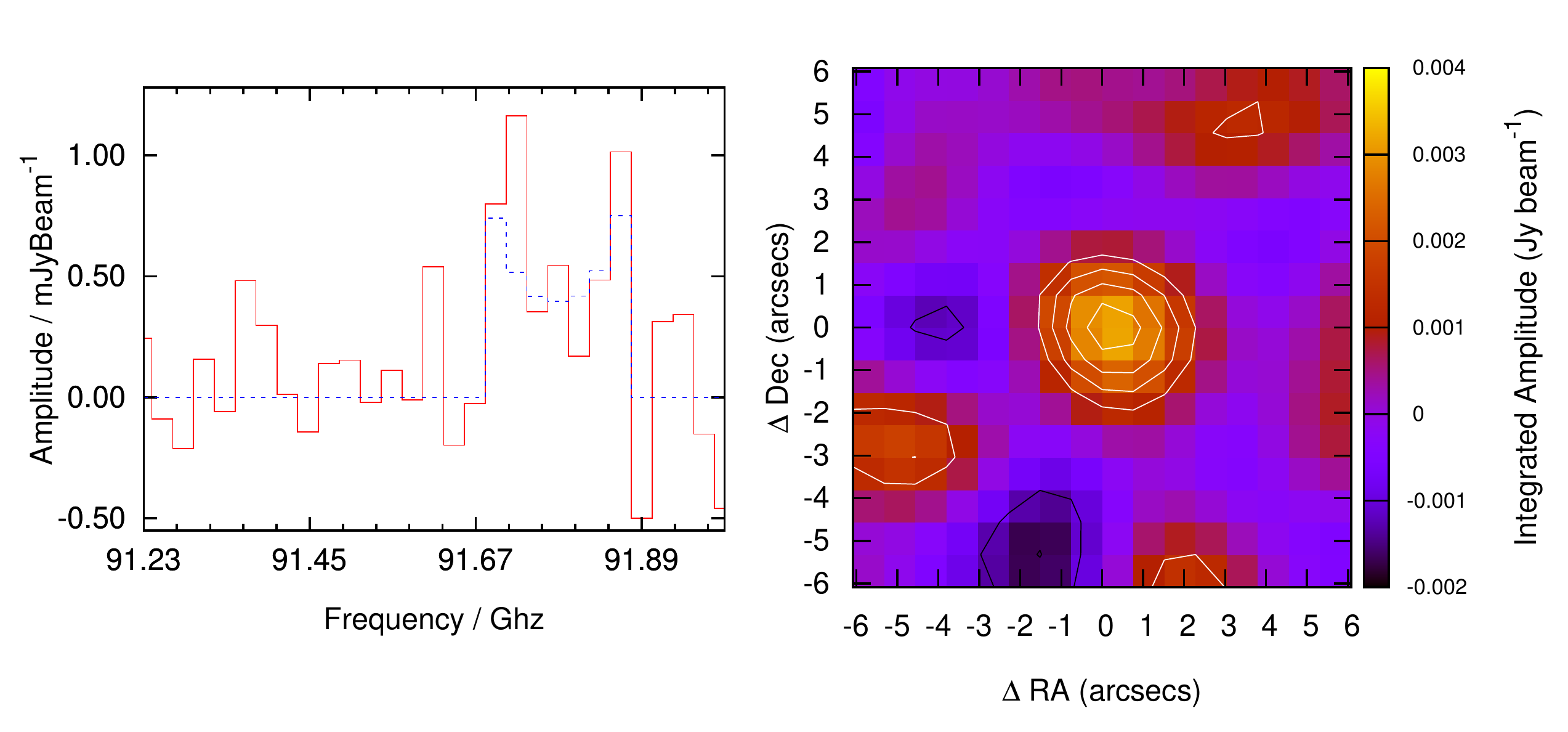} \\
\hspace{-1.5cm}
\includegraphics[width=100mm]{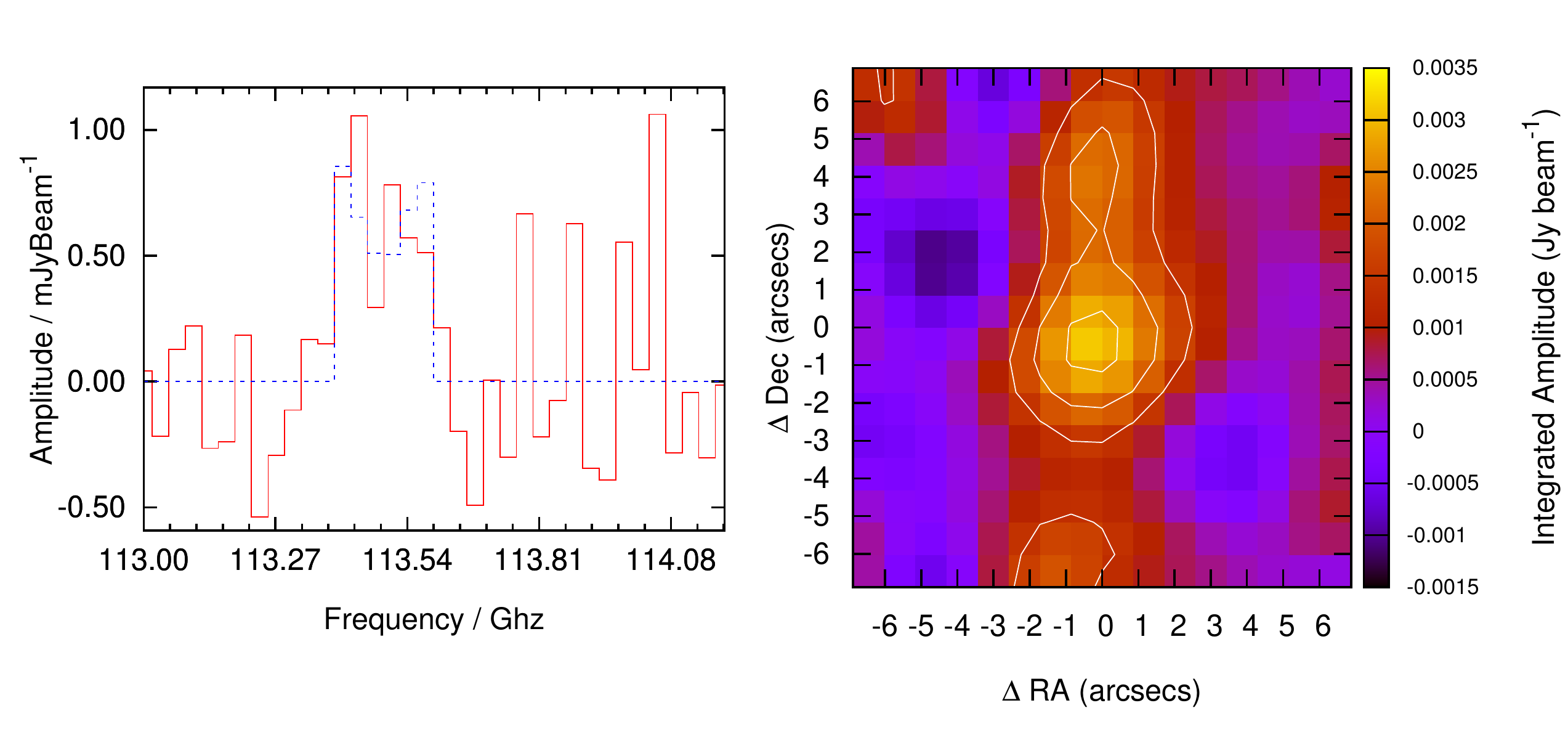} &
\includegraphics[width=100mm]{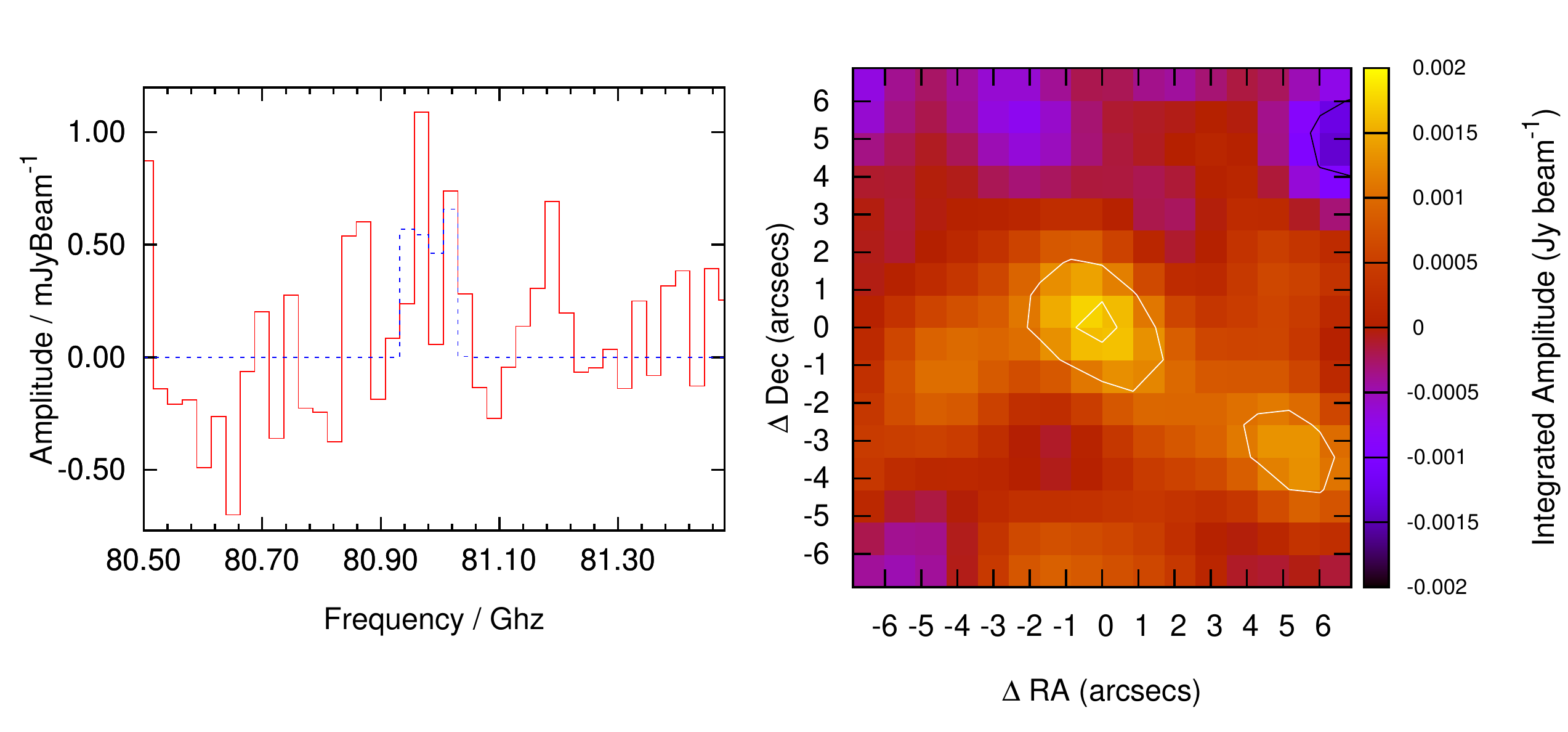} \\
\end{array}$
\end{center}
\caption{Maps and spectra for (top) sources Sh1-1 and Sh2-1 from Table \ref{Table:MainResults} and  (bottom) the 2 highest probability candidate detections from the directed search, Z1 and Z2, from Table \ref{Table:specsourcelist}. Spectra and maps are centered at the nominal location of the candidate given in Tables \ref{Table:MainResults} and \ref{Table:specsourcelist}, unless they are close to the edge of a spectral window in which case they will appear as slightly offset.  Spectra show the data (red solid line) with the best fit spectral model overlaid (blue dotted line).  Maps are produced by integrating over those channels in which the model amplitude of the source is greater than twice the noise level in the image, with units in Jy~beam$^{-1}$.   Postive contours are shown in white, negative contours in black, with initial contours at $\pm$2$\sigma$ increasing by 1 each time.}
\label{figure:redshiftimages}
\end{figure*}

In order to assess the Gaussianity of the noise in the survey we also perform our search twice more, each time randomising the order of the channels.  We denote detections in these cubes as Sh1-$X$ and Sh2-$X$ for each reordering respectively. Spectra and maps for the two most significant of these detections are shown in Fig \ref{figure:redshiftimages} (top).

Assuming an emission line with FWHM of 300 km~$\mathrm{s^{-1}}$, and given a search area in each channel of approximately 4000 arcseconds$^2$ and a resolution of 9 arcseconds$^2$ $\sim$ 160000 independant samples will be present in the search volume.  Given simple Gaussian statistics we would therefore expect to find $\sim$ 1 $4.5\sigma$ detection in our analysis which we would expect to assign a probability of $\sim 50\%$.  

We note however that, when randomising the channels, Table \ref{Table:MainResults} lists $\sim 4-5$ candidate detections at a significance of 4.5$\sigma$ and above for both cases despite the fact that this should have been representative of purely noise in the data.  This is highly suggestive of non-Gaussian behaviour in the noise, which is to be expected given the sparse UV coverage of the PdBI scan.  We can attempt to account for this in a crude way by considering that, if we detect $\sim$ 4 times as many noise candidates as we expect, that is akin to searching in a volume 4 times larger.  As such we can rescale $\mu_{\mathrm{s}}$, and subtract $\log 4$ from our evidence values in order to estimate new probabilities for our candidate detections.  We include this scaled probability, which we will denote $\hat{\mathrm{P}}$, in Table \ref{Table:MainResults}.

Comparing the number of detections after rescaling our probabilities in the two randomised datasests with the actual survey, we find one detection above our acceptance threshold compared to 4 detections in the survey. As we increase $\mu_{\mathrm{s}}$ the probabilities of detections in the randomised datasets are similar to those in the real dataset suggesting that the reliabiity of any detections below this point will be low.  In constructing a final list of sources we therefore only consider the four candidates with $\hat{\mathrm{P}}>50\%$ for the case of $\mu_{\mathrm{s}}=3$, brief details of the individual sources will be given below, with reference to D14 for more details where the candidates are present in both catalogues.

\begin{table*}
\caption{Model Parameters for galaxy candidates in the HDF blind survey with probabilities greater than 50$\%$ when assuming different values for $\mu_{\mathrm{s}}$.  For each source we list the: i) the value of $\mu_{\mathrm{s}}$ used to calculate the probability, ii) the source ID, where a candidate corresponds to a source ID in D14 we also list the corresponding identifier [ID$X$], we also list the parameters for the candidate detections when randomising the channels which we denote Sh1-$X$ and Sh2-$X$ for each reordering respectively, iii) best fit positions and frequencies, iv) FWHM, defined to be either 2.3548$\sigma_v$ when using a Gaussian model, or 2$\sigma_v$ for the Brandt paramaterisation, v) the primary beam corrected velocity integrated CO flux ($I_{\mathrm{CO}}$), vi) S:N ratio of the source defined in terms of Eq. \ref{Eq:ImProbFinal} such that S:N=$\hat{A} /\hat{\sigma}$, where the kernal used for the weighted sum is the mean source model returned for each candidate, vii) candidate probability, and finally viii) rescaled probability. Spectra and maps are shown for each of the sources where $\mu_{\mathrm{s}}=3$ in Fig. \ref{figure:blind}.}
\centering
\begin{tabular}{ccccccccccc}
\hline\hline
$\mu_{\mathrm{s}}$ & ID 	& RA 	& 	DEC	&	Frequency	& 	FWHM &  $I_{\mathrm{CO}}$ 		&   S:N	& P	& $\hat{\mathrm{P}}$ &\\ [0.5ex]
      		& 		&J2000.0&J2000.0	&          Ghz		&	km~$\mathrm{s^{-1}}$ & Jy~km~$\mathrm{s^{-1}}$		&	&	&	&\\
\hline
3 & B1 [ID3] &	12  36  48.2724  &  62  12  15.9273   &   82.788  $\pm$ 0.003 & 341 $\pm$ 30 & 0.49 $\pm$ 0.03 &  7.4 &  100 &	99.6	\\
3 & B2 [ID17] &	12  36  51.7762  &  62  12  26.2030   &   111.852 $\pm$ 0.03 & 540 $\pm$ 71 & 0.48 $\pm$ 0.08 &  4.9 &  95.7 &	84.8	\\
3 & B3 [ID18] &	12  36  48.6803  &  62  12  38.8353   &   112.621 $\pm$ 0.013 & 350  $\pm$ 90& 0.56 $\pm$ 0.12 &  5.4 &  93.1 &	77.1	\\
3 & B4 	&	12  36  51.9045  &  62  12  14.9837   &   85.237 $\pm$  0.006 & 650 $\pm$ 40 & 0.38 $\pm$ 0.06 &  5.3 &  88.4 &	65.6	\\
3 & Sh1-1	&	12  36  48.9486  &  62  12  56.5618   &   81.657 $\pm$  0.008 & 350 $\pm$ 110 & 0.60 $\pm$ 0.11 &  4.8 &  81.4 &	52.3	\\ 	
3 & B5 	& 	12  36  50.1098  &  62  12  50.3014   &   104.835 $\pm$ 0.03 & 420 $\pm$ 140 & 0.65 $\pm$ 0.16 &  4.5 &  58.9 &	26.4	\\
3 & B6 	& 	12  36  49.8638  &  62  12  33.8323   &   111.338 $\pm$ 0.012 & 500 $\pm$ 70 & 0.37 $\pm$ 0.07 &  5.8 &  53.0 &	22.0	\\
3 &  Sh2-1     &	 12  36  52.7266  &  62  12  56.9566  &   91.779  $\pm$  0.017 & 600 $\pm$ 100 &  1.3 $\pm$ 0.3 &  5.4  & 51.9 &	21.2	\\

4 &  Sh2-2     &	 12  36  49.1675  &  62  12  22.7874  &  107.784 $\pm$ 0.016 & 420 $\pm$ 110  & 0.37  $\pm$ 0.08 &  4.5  & 54.5 &	23.0	\\
4 &  Sh2-3     &	  12  36  50.1260  &  62  12   2.8105  &   92.900   $\pm$ 0.017 &130 $\pm$ 20 & 0.56 $\pm$ 0.14 &  5.0  & 51.9 &	21.2	\\
4 & B7 [ID1]  &	12  36  47.5851  &  62  12  19.8882   &  80.048  $\pm$ 0.02 & 590 $\pm$ 110 & 0.45 $\pm$ 0.10 &  5.1 & 51.9&	21.2	\\
4 &  Sh2-4     &	 12  36  48.5616   & 62  12   2.7008   &  91.227 $\pm$ 0.03 & 500 $\pm$ 110 & 0.62 $\pm$ 0.14 &  4.9  & 51.7 &	21.1	\\

5 & B8 [ID8] & 	12  36  51.9270  &  62  12  26.6436   &   93.171 $\pm$ 0.013 & 420 $\pm$ 130 &  0.35 $\pm$ 0.08 &  5.1 &  54.2 &	22.8	\\
5 &  Sh1-2     &	 12  36  49.3257 &   62  13   2.3787  &   86.264 $\pm$ 0.027 & 580 $\pm$ 65 & 0.98 $\pm$ 0.18 &  4.5  & 51.4 &	20.9	\\
5 &  Sh1-3     &	  12  36  48.9928  &  62  12  42.4842  &   93.002 $\pm$ 0.006 & 340 $\pm$  60 & 0.47 $\pm$ 0.07 &  4.7  & 51.2 &	20.8	\\

5 & B9 [ID5]    &	12  36  52.3523  &  62  12  11.3842   &   89.879 $\pm$  0.03 & 420 $\pm$ 130 & 0.44 $\pm$ 0.1 &  4.7 & 54.0 &	22.7	\\
5 &  Sh2-5     &	 12  36  50.3969  &  62  12  23.3128   &  87.710  $\pm$ 0.012  & 410 $\pm$ 110 & 0.29 $\pm$ 0.06 &  5.3  & 54.0 &	22.7	\\

6 & B10 &		12  36  50.2527  &  62  12  12.2119   &   97.505 $\pm$ 0.02 & 500 $\pm$ 110 & 0.42 $\pm$ 0.10 &  4.2 & 52.1 &	21.4	\\
6 &  Sh1-4     &	 12  36  49.5080   & 62  12  29.8296  &  103.991  $\pm$ 0.014 & 520 $\pm$ 70 & 0.34 $\pm$ 0.05  &  5.4  & 51.1 &	20.7	\\

\hline
\end{tabular}
\label{Table:MainResults}
\end{table*}

\begin{itemize}
\item{B1} Corresponds to ID3 in D14. It is coincident with an optically faint but infrared bright galaxy with AB magnitudes that follow the definition of a BzK galaxy (see D14 for details) and photometric redshift of $z_{\mathrm{phot}} \sim 1.6$, close to the CO redshift of $z=1.784$ for a CO($2\to1$) transition.\\

\item{B2} Corresponds to ID17 in D14.  Identified as the CO($6\to 5$) transition for the SMG galaxy HDF850.1 at $z=5.183$.  The second expected emission line corresponding to the CO($5\to 4$) transition corresponds to candidate B8 in Table \ref{Table:MainResults} but falls below the threshold for acceptance in the context of a blind survey. We shall show in section \ref{Section:Directed} that when including additional prior information in the form of position and redshifts of known sources that this probability increases to be a definite detection in the data.  Refer to \cite{2012Natur.486..233W} for more details on HDF850.1.\\

\item{B3} Corresponds to ID18 in D14.  No galaxy counterpart observed at optical/NIR/MIR wavelengths, most likely corresponds to the CO($3\to2$) transition at $z=2.071$.\\

\item{B4} Has no counterpart in the D14 catalogue.  There are two possible counterparts within 1.5$\arcsec$ in the photometric redshift catalogue \cite{1999ApJ...513...34F} (henceforth FLY99) with $z_{\mathrm{phot}} \sim 3.92$ and $z_{\mathrm{phot}} \sim 0.48$.  Given the quoted uncertainties of $\pm 0.1~(1+z_{\mathrm{phot}})$ this could therefore correspond to either ID within 1$\sigma$ uncertainties, corresponding to either the CO($3\to2$) or CO($4\to3$) transition in the first case or the CO($1\to0$) transition in the second. Given the volumes searched at these relative redshifts the $z_{\mathrm{phot}} \sim 3.92$ counterpart is the more probable, with the CO($4\to3$) transition closer to the frequency of the line in our data.\\

\end{itemize}

\subsection{Directed Searches}
\label{Section:Directed}

\begin{table*}
\begin{minipage}{126mm}
\caption{Positions and redshifts of galaxies with spectroscopic or grism-based redshifts used in the directed search for molecular CO.}
\centering
\begin{tabular}{cccccccccc}
\hline\hline
ID 	& RA 		& 	DEC		&	$z_{\mathrm{spec}}$	& 	$z_{\mathrm{grism}}$ & $\Delta z$ & Transition	& P & $\hat{\mathrm{P}}$	\\ [0.5ex]
	& 	J2000.0 	&	J2000.0	&				&			    &			&			& &	\\
\hline
Z1 [ID19]	& 12:36:51.61   & +62:12:17.3 & 		 		&	2.044	    &    0.015	& CO($3\to2$)			&	99.8	&99.2	\\
Z2	& 12:36:47.28   & +62:12:30.7 & 	0.4233 		&			    &    2E-4	& CO($1\to0$)				&	76.46	&44.8	\\
Z3	& 12:36:52.67   & +62:12:19.8 & 	0.401 		&			    &    2E-4	& CO($1\to0$)				&	48.7	&	19.2\\
Z4	& 12:36:49.56   & +62:12:36.1 & 		 		&	2.014	    &    0.015	& CO($3\to2$)				&	46.42	&17.8	\\
Z5	& 12:36:52.09   & +62:12:26.3 & 	1.224 		&			    &    3E-4	& CO($2\to1$)				&	44.88	&16.9	\\
Z6	& 12:36:53.49   & +62:12:31.7 & 		 		&	1.125	    &    0.011	& CO($2\to1$)				&	36.14	&12.4	\\
Z7	& 12:36:46.24   & +62:12:29.1 & 		 		&	1.585	    &    0.013	& CO($2\to1$)				&	29.3	&9.4	\\
Z8	& 12:36:49.99   & +62:12:26.3 & 		 		&	1.284	    &    0.011	& CO($2\to1$)				&	25.3	&7.8	\\
Z9	& 12:36:46.94   & +62:12:26.1 & 	2.97	 		&			    &    0.005	& CO($3\to2$)			&	21.72	&6.5	\\
Z10	& 12:36:51.28   & +62:12:33.8 & 		 		&	1.862	    &    0.014	& CO($2\to1$)				&	21.25	&6.3	\\
Z11	& 12:36:49.81   & +62:12:48.8 & 	3.233 		&			    &    0.006	& CO($3\to2$)				&	15.5	&4.4	\\
Z12	& 12:36:50.48   & +62:12:50.4 & 	4.345 		&			    &    0.008	& CO($4\to3$)				&	15.35	&4.3	\\
Z13	& 12:36:47.49   & +62:12:11.2 & 		 		&	1.58		    &    0.013	& CO($2\to1$)			&	15.01	&4.2	\\
Z14	& 12:36:53.42   & +62:12:21.7 & 		 		&	1.715	    &    0.014	& CO($2\to1$)				&	13.42	&3.7	\\
Z15	& 12:36:46.22   & +62:12:28.5 & 		 		&	1.591	    &    0.013	& CO($2\to1$)				&	12.1	&3.3	\\
Z16	& 12:36:49.81   & +62:12:48.8 & 	3.233 		&			    &    0.006	& CO($4\to3$)				&	7.8	&2.1	\\
Z17	& 12:36:47.04   & +62:12:36.9 & 	0.3209 		&			    &    2E-4	& CO($1\to0$)				&	4.72	&1.2	\\
Z18	& 12:36:49.95   & +62:12:25.5 & 	1.204 		&			    &    3E-4	& CO($2\to1$)				&	5.82	&1.5	\\
Z19	& 12:36:53.66   & +62:12:23.7 & 	1.731 		&			    &    0.003	& CO($2\to1$)				&	5.3	&1.4	\\
Z20	& 12:36:51.71   & +62:12:20.2 & 	0.3			&		    &   2E-4	& CO($1\to0$)				&	3.18	&0.8	\\
\hline
\end{tabular}
\label{Table:specsourcelist}
\end{minipage}
\end{table*}

One of the advantages of taking a Bayesian approach is that, when additional prior information becomes available, that information can be folded into the analysis, and the posterior parameter estimates and probabilities can be re-evaluated.  In the context of source finding, if the position or redshift of a source is known, we can include that information in our priors when performing the search.  In doing so, a source candidate that might have a low probability in the context of a blind survey could have much higher probability in a directed search.  

Table \ref{Table:specsourcelist} gives details of a set of 20 galaxies with high quality known red shifts as discussed in W14.  In brief, these redshifts are based on either long-slit spectroscopy from the Keck telescope, or grism-based redshifts based on the detection of emission lines from the HST survey 'A Grism H-Alpha SpecTroscopic survey'   (AGHAST, Weiner et al 2013 in prep).  Uncertainties for grism-based redshifts are taken to be $\Delta z = 0.005(1 + z)$, whilst the spectroscopic redshifts have typical uncertainties of $\sim$ 50 km s$^{-1}$ for $z \le 1.6$, whilst at higher redshifts these uncertainties increase to $\sim$ 400 km s$^{-1}$.

Priors on redshift for each transition are listed in Table \ref{Table:specsourcelist}, we consider a search area $\pm 1.5\arcsec$ from the nominal position quoted, and for any remaining parameters we use the same priors described in section \ref{Section:Blind}.  We then performed the directed search for each emission line using the three spectral models listed in Section \ref{Section:SpecModel}.  The highest probability associated with these different models for each emission line is given in Table \ref{Table:specsourcelist}.

Applying the same rescaling of probabilities to the results of the directed search as in the blind survey we find only Z1 is a significant detection in the data with a probability of 99$\%$.  Z1 corresponds to ID19 in D14, the second source assigned a secure rating in D14 which we therefore confirm with the directed Bayesian analysis.  One can immediately see the benefit of including the additional position and redshift information in the priors for the search, as in the context of a blind survey, Z1 had a probability of less than $50\%$ and thus was not significant enough a detection to warrant inclusion in the final list of candidates.   

In W14 Z14 is also identified as a tentative detection of the CO($4\to3$) line, however we find no support for such a claim in our analysis, with a probability of only 4$\%$.

\subsection{Comparison to D14}

\begin{table}
\caption{Probabilities for the candidate sources in D14}
\centering
\begin{tabular}{cccc}
\hline\hline
D14 ID 	  &  D14 Quality & SN$_{\mathrm{spec}}$ & $\hat{\mathrm{P}}$ \\ [0.5ex]
          & 		 & 	     &	 \\
\hline
1  &   	2		&4.6	&  17.0\\
2  &	2		&3.5	&  0.7\\
3  &	1  - secure 	&6.3 	& 99.6 \\
4  &    3		&2.7	&  0.1\\
5  &   	2		&4.2	&  10.3\\
8  &	2  - HDF850.1	&4.4	&  17.0\\
10 &	2   		&4.2	&  0.6\\
11 &    2		&3.6	&  1.1\\
12 &   	2		&3.5	&  1.4\\
13 &	3		&2.1	&  0.3\\
14 &	3   		&3.4	&  11.9\\
15 &    2		&4.1	&  1.2\\
17 &   	2 - HDF850.1	&3.6	&  84.8\\
18 &	1		&5.1	&  77.1\\
19 &	2 - secure	&3.8	&  99.2\\
20 &    2		&3.6	&  4.7\\
21 &    3		&2.7	&  0.3\\
\hline
\end{tabular}
\label{Table:DComp}
\end{table}

In D14 candidate detections are assigned a quality rating from 1-3 using a spectrum-based S:N, which we will denote SN$_{\mathrm{spec}}$, computed as the ratio between the fitted Gaussian line flux and its uncertainty. The ratings then correspond to: quality flag 1) objects with SN$_{\mathrm{spec}} > 5$, quality flag 2) those with SN$_{\mathrm{spec}}$ between 3.5 and 5, and quality flag 3) those with SN$_{\mathrm{spec}} < 3.5$.  Two candidates are, in addition to this quality flag, assigned a `secure' rating, corresponding to those candidates which have been confirmed through follow-up observations.

Table \ref{Table:DComp} lists the probabilities of the 17 candidate sources from D14 that result from either the blind search described in section \ref{Section:Blind}  with $\mu_{\mathrm{s}} = 3$ for all sources, or the directed search in section \ref{Section:Directed}.  In addition to these D14 includes 4 sources with negative fluxes which we have not listed in this table.  

Of the two detections given a secure rating in D14 one (B1) is detected in our blind analysis at high significance, while the second was highly significant in the context of the directed search, in addition, both candidates assigned a quality rating of 1 are returned with high probability in the Bayesian analysis.

In comparison, of the 11 candidates given a quality rating of 2, only one of the two lines that correspond to HDF850.1, and the second secure detection are found to exceed our threshold for acceptance of $50\%$.  The remaining quality 2, and 3 candidates are all $\sim$ 4.5$\sigma$ and below which as discussed in Section \ref{Section:Blind} means that given the non-Gaussian nature of the noise in the dataset they are unlikely to have significant probabilities.

\section{Conclusions}
\label{section:conclusions}

We have presented an efficient and statistically robust Bayesian approach to detecting galactic emission lines in the image domain.  In the most general case we are able to parameterise both the spatial and spectral characteristics of each of the detected galaxies during the search, whilst using the Bayesian evidence to return the probability that any given detection is 'real'.  When detections are made the evidence can also be used to select between different source models in order to find the optimal description supported by the data.  

Using observations taken with the Plateau de Bure Interferometer of the Hubble Deep Field North over the entire 3mm window (79.7-114.8 GHz) we have used this technique to perform the first Bayesian blind survey for cold molecular gas in the Universe. In the context of a blind survey, where no additional prior information is included in the search such as known source positions or redshifts a total of 4 detections are made with probabilites that exceed our threshold for acceptance:

\begin{itemize}
	\item The most significant source is associated with the CO($2 \to 1$) transition for a BzK galaxy at redshift 1.784.  Followup observations presented in D14 confirm the detection.
  	\item The second most significant line is associated with the CO($6\to5$) transition line for the SMG HDF850.1.
	\item The third most significant has no known counterpart in any optical/NIR/MIR wavelengths
	\item The final detection has two possible counterparts in the FLY99 photometric redshift catalogue within 1.5$\arcsec$ with redshifts that are within $1\sigma$ of the CO detection.  The most likely of these has the detection represent the CO($4\to3$) transition at a redshift of $z= 4.41$.
\end{itemize}
We then perform a directed search - including position and constraining redshift information for a set of 20 known sources in the field.  In doing so we are able to probe lower signal to noise levels and detect a further source: 

\begin{itemize}
	\item A BzK galaxy at a redshift of $z=2.048$ with an optical/NIR counterpart 
\end{itemize}
This study acts as a demonstration for how to perform a robust statistical analysis of blind surveys of molecular gas in the Universe.  In the near future surveys will be carried out both for low-$J$ CO transitions of high redshift galaxies using the JVLA, and at millimeter wavelengths with ALMA.  Here the sensitivies reached will allow us to explore currently inaccessible portions of the CO luminosity function, and an analysis such as that presented here will be required in order to exract the most from the data, inferring reliable scientific conclusions about the history of the molecular gas properties of star-forming galaxies in the Universe through cosmic time.

\section*{Acknowledgments}

\appendix

\bsp

\label{lastpage}

\end{document}